A transition of dynamo modes in M dwarfs: narrowing down the spectral range where the transition occurs[*]

D. J. Mullan[1] and E. R. Houdebine[2]

[1]Dept of Physics and Astronomy, University of Delaware, Newark DE 19716 USA

[2]Armagh Observatory, College Hill, BT61 9DG Armagh, UK; Université Paul Sabatier, Observatoire Midi-Pyrénées, CNRS, CNES, IRAP, F-31400 Toulouse, France

**Abstract**

Houdebine et al (2017: H17) combined CaII data with projected rotational velocities (v sin i) to construct rotation-activity correlations (RAC) in K-M dwarfs. The RAC slopes were used to argue that a transition between dynamo modes occurs at a spectral type between M2 and M3. H17 suggested that the dynamo transition corresponds to a transition to complete convection (TTCC). An independent study of GAIA data led Jao et al (2018) to suggest that the TTCC sets in near M3.0V, close to the H17 result. However, the changes in a star which cause TTCC signatures in GAIA data might not lead to changes in CaII emission at an identical spectral type: the latter are also affected by magnetic effects which depend on certain properties of convection in the core. Here, we use CaII emission fluxes in a sample of ~600 M dwarfs, and attempt to narrow down the transition from one dynamo mode to another: rather than relying on RAC slopes, we quantify how the CaII emission flux varies with spectral type to identify steps where the flux decreases significantly across a narrow range of spectral types. We suggest that the dynamo mode transition may be narrowed down to between M2.1 and M2.3. This is close to, but earlier than, the TTCC location identified by Jao et al (2018). We suggest that the transition in dynamo mode may be related to the existence of a small convective core which occurs for a finite time interval in certain low mass stars.

Key words: stars: chromospheres; stars: acoustic fluxes; stars: magnetic activity

## 1. Introduction

Houdebine et al (2017) (hereafter referred to as H17) reported on measurements of the fluxes of chromospheric emission in the CaII lines in a sample of (272) lower main sequence stars with spectral sub-types of K4, K6, M2, M3, and M4. Using archival spectra which had been obtained over many years by the HARPS and SOPHIE spectrometers, the flux F(CaII) of the CaII emission in each star was extracted.

It is important to clarify how, in the present paper, we define "the" flux F(CaII) for each star in our sample. It is known that chromospheres in stars are variable on time-scales which range from as short as days (rotational modulation) to as long as years (activity cycles?) (e.g. Suarez Mascareno et al 2016). The spectra we analyze in the present paper were derived from archival data which were gathered originally in connection with planet search programs using the SOPHIE, HARPS and FEROS instruments. These observations, designed to search for small variations in radial velocity associated with reflex motion of the star as a planet orbits, are distributed over a span of many years. As a result, the archival data include many spectra for each star, distributed randomly in time. Typically, each star in our sample has dozens of spectra: many have 100 or more, and the limiting case is more than 700 for Prox Cen. Stars that were observed a few times only are in the minority. Therefore, when we report "the" F(CaII) for any particular star, the numerical value is an average over multiple random epochs. The averaging process smooths out any variations due to rotation and/or activity cycles.

An anonymous referee has raised the issue of possible misclassifications of active/inactive stars due to variability. In order to avoid such difficulties, we have excluded from our statistics the subdwarfs and the intermediate dM(e) stars which have peculiar metal abundances and/or spectral properties. Only normal low activity and high activity dwarfs are included in our samples.

As a result of these precautions, our measured EWs can be considered as representative averages of "the" CaII emission flux for each star.

### 1.1. Chromospheric heating and dynamo operation

The flux of CaII emission from a star is a (partial) measure of how effective the chromospheric heating is in that star. In any star, mechanical work of some kind is required to heat the chromosphere to temperatures which exceed the photospheric value: the transport of heat inside the star which leads to a temperature that declines with increasing radial location is a natural feature of radiative and convective transfer of energy outward. But the onset of increasing local temperatures above a certain height requires the presence of an agent (or agents) which perform work on the local material in order to raise the temperature a few thousands of degrees above that predicted by radiative equilibrium (e.g. Mullan 2009). Agents which perform such work include acoustic waves and also magnetic fields. All low-mass stars possess deep convective envelopes where acoustic waves are inevitably generated by the ubiquitous pressure fluctuations that are inevitable in a compressible medium: as a result, acoustically-heated chromospheres are present in all low-mass stars. But stars can also generate chromospheres by means of heating due to magnetic processes, including dissipation of MHD waves of various kinds (Alfvenic, slow MHD, fast MHD) (e.g. Osterbrock, 1961), dissipation of electric currents (e.g. Goodman 1995), and nanoflares (e.g. Jess et al 2014). Naturally, the magnetic mechanisms require the presence of magnetic fields in the star. Schrijver (1983) suggested that at any given spectral type, stars with the weakest chromospheric emissions might be regarded as representatives of the lowest permissible heating: the term "basal flux" was coined to refer to the lower limit on mechanical energy flux in a star of any particular spectral type. It is widely believed that the chromosphere in such stars is acoustically heated. Stars with chromospheric emissions which exceed the basal fluxes are considered to contain sources of both acoustic plus magnetic heating. In the present paper, we are interested in stars where the chromospheres lie close to the basal flux, but which also include at least some magnetic contributions.

As regards the magnetic contributions, dynamos of three major kinds have been modeled in the literature (e.g. Racine et al 2011): these are referred to by the labels $\alpha\Omega$, $\alpha^2$, and $\alpha^2\Omega$. In these labels, the parameter $\alpha$ quantifies a physical process (kinetic helicity) which, by means of local motions of the gas, creates an electric field that is related to the strength of the mean magnetic field. Parker (1955) suggested that the "$\alpha$-effect" could occur in convective turbulence if cyclonic motions (driven by rotation) were available to systematically deflect and twist the stellar magnetic field. Also in the above labels, $\Omega$ is the angular velocity: more formally, the important physical parameter in dynamo operation is not so much the absolute magnitude of $\Omega$, but rather the radial *gradient* of $\Omega$.

It is important, in the context of the present paper, to note that some of the stars in our sample are sufficiently massive that they contain an interface between a radiative core and a convective envelope inside the star. At such an interface, physical conditions may lead to a steep local radial gradient of $\Omega$ inside a thin turbulent boundary layer: this layer (the tachocline) is expected to be the site of effective dynamo activity (e.g. Spiegel and Zahn 1992). In fact, as regards dynamo activity in our own Sun, arguments can be made that the site of solar activity in fact lies in a layer *near the lower edge* of the tachocline (Stenflo 1991). Stars on the lower main sequence are expected to have convective envelopes which may be considerably deeper than the zone in the Sun (Stromgren 1952; Osterbrock 1953): in limiting cases, the convection zone may extend all the way to the center of the star. Based on theoretical work by Limber (1958), who computed the first fully convective models of M dwarfs, it is expected that in some of the stars in our sample, a tachocline is indeed present, whereas in other stars in our sample, there is no such interface. In the latter case, the star is completely convective. According to a particular model of stellar structure, the transition to complete convection (TTCC) is predicted to occur among main-sequence stars with masses in the range 0.32-0.34 $M_\odot$ (Mullan et al 2015).

It is important to note that the overall sensitivity to rotation varies from one kind of dynamo process to another. These variations play an essential role in the interpretation we place on chromospheric heating in low-mass stars.

Observational evidence based on Zeeman-Doppler imaging of low-mass stars indicates that certain differences can be identified between the magnetic field properties in stars above and below the TTCC. E.g., the surface fields tend to be stronger in certain fully convective stars than in stars with a radiative core, although weak fields can also be present on other fully convective stars (Morin 2012). Moreover, completely convective stars generate fields which are predominantly poloidal, whereas in stars above the TTCC (i.e. stars with radiative cores), the fields are predominantly toroidal (See et al 2016). Our goal is to determine if any observational signature can be identified in CaII emission data associated with the TTCC.

The plan of the paper is as follows. In Section 1.2, we describe how the rotational sensitivity of CaII emission in K and M dwarfs can be quantified: specifically, we examine the empirical rotation-activity correlation (RAC), and extract numerical values for the *slope* of the RAC. The results which emerged (in H17) from our use of the RAC slope technique are summarized in Section 1.3. Our interpretation of these results in terms of dynamo models are summarized in Section 1.4, where we recapitulate the major conclusion of H17: something interesting happens in the CaII emission properties of stars with spectral types between M2 and M3, perhaps indicating a change in the dynamo mode. In Section 1.5, we raise the question: can more than one dynamo mechanism be at work in a given star? In Section 1.6, we summarize a completely independent study (using GAIA data) which, more than a year after the H17 paper appeared, reported that the stellar luminosity function exhibits an empirical feature "near spectral type M3.0V". We summarize theoretical work which attributes the empirical feature to the onset of complete convection, and discuss how this result provides independent support for the major conclusion of H17. In Section 1.7, we turn to consider the information which was contained in our earlier paper (H17) as regards the *absolute levels* of the fluxes of chromospheric emission in CaII as a function of spectral type. The main purpose of the present paper is to extend the H17 data set with new data on CaII emission (in terms of equivalent widths and absolute fluxes) in a larger sample of stars. These new data are presented in Section 2, where we show (in Figs. 3, 4, and 5) how the CaII emission flux declines as we go towards later spectral sub-types among inactive and active M dwarfs. In Section 3, our CaII empirical emission fluxes in inactive M dwarfs are compared with theoretical predictions which have been made by Ulmschneider et al (1996) for the acoustic power generation in cool dwarfs. In Section 4, we raise the issue: does any significant "feature" exist in the curve of CaII emission fluxes versus spectral type? We discuss if such a "feature" might be associated with a change in dynamo behavior when a low-mass main-sequence star becomes completely convective. Conclusions are in Section 5.

**1.2. Distinguishing between different processes of chromospheric heating: the RAC slope technique**

Is there a way to distinguish between a chromosphere in a star where one form of heating dominates from the chromosphere of another star in which another heating mechanism dominates? One possibility is to examine the rotation of the star: dynamos are expected to rely to a greater or lesser extent on rotational motion in order to operate, whereas acoustic fluxes are not expected to be sensitive to rotation. Thus, if we can identify an empirical relationship between rotation and magnetic activity (the latter being quantified by the strength of chromospheric emission), separation of one form of heating from another might be possible. This applies not merely to distinguishing acoustic heating from magnetic heating: we might also hope to distinguish one type of magnetic heating from another.

In H17, the fluxes of emission in the chromospheric lines of CaII were obtained from archival spectra. But also from the same spectra, rotational information (v sin i) was obtained by cross-correlating the profiles of hundreds of photospheric lines. By plotting F(CaII) emission fluxes *versus* the (projected) rotational period P/sin i for our sample of stars in each spectral sub-type, a rotation-activity correlation (RAC) was derived for each of the 5 spectral sub-types. For stars in each spectral sub-type, in a plot of the RAC, i.e. in a plot of log[F(CaII)] *versus* log[P/sin i], least-squares fitting was used to determine a value for the slope of the best-fitting line. In all cases, the slope was found to have a negative value, i.e. the faster a star rotates, the stronger is the chromospheric emission in CaII. The chromospheric emission is sensitive to rotation.

The absolute magnitude of the slope which is associated with the RAC at any particular spectral sub-type is a measure of how sensitive the chromospheric emission is to rotation for stars of that spectral sub-type. For stars of some spectral sub-types, the slope is found to have a large absolute magnitude: for such stars, we say that the RAC has a *steep* slope. If a steep RAC slope is discovered in any sample of stars, this suggests the operation of a dynamo in which rotation plays a dominant role. In such stars, it is believed that there is a dynamo at work such that faster rotation generates significantly stronger magnetic fields: in the presence of such stronger magnetic fields, stronger chromospheric heating is expected, thereby giving rise to more intense emission in (e.g.) the CaII lines. The dynamo in such stars, where rotational sensitivity is maximal, might be an interface dynamo, perhaps an $\alpha\Omega$ dynamo.

On the other hand, if a *shallow* RAC slope is discovered in a sample of stars, this suggests that rotation plays a less important role in the dynamo operation. An example of such a dynamo is an $\alpha^2$ (or $\alpha^2\Omega$) dynamo, which relies on the existence of small-scale turbulent eddies distributed throughout the extensive convective envelope of a low-mass star. The dependence on eddies with a wide spatial distribution leads to terms such as "distributed dynamo" or "small-scale dynamo": in such a dynamo, rotation still contributes somewhat to the dynamo process, but the contribution is not as dominant as in an $\alpha\Omega$ dynamo (Durney et al 1993). In what follows, we shall use the short-hand notation $\alpha^2$ dynamo as a proxy for a distributed or small-scale dynamo in which the effects of rotation are relatively weak.

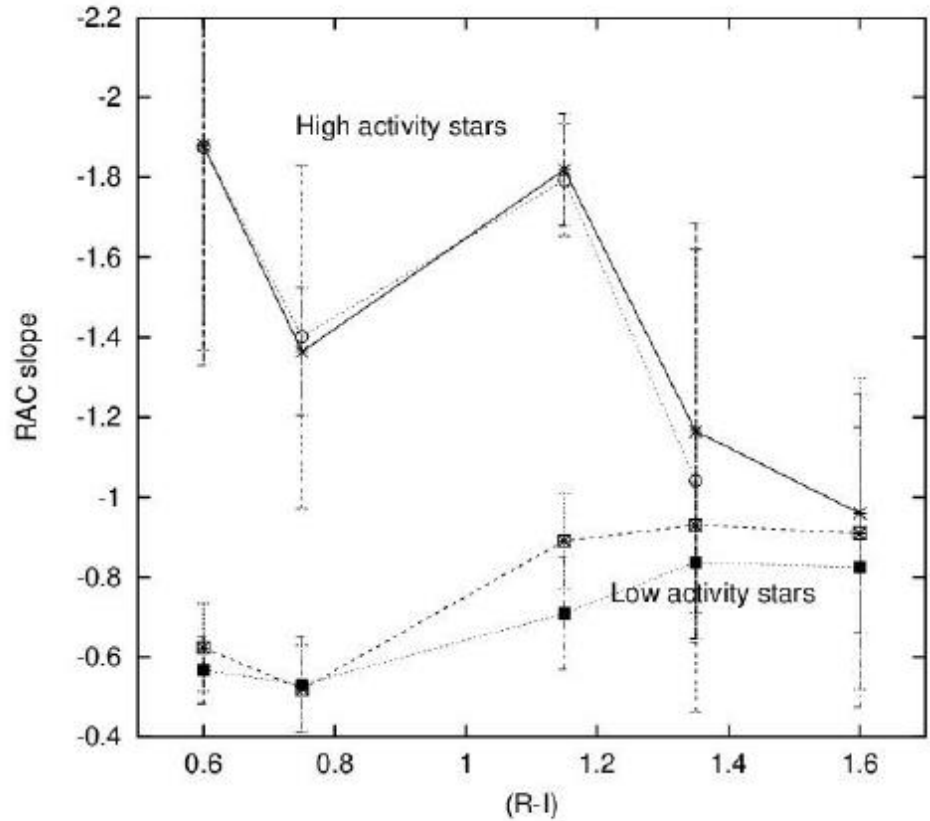

Figure 1. Numerical value of the **slope** of the rotation-activity correlation (RAC) for samples of stars belonging to different spectral sub-types. Each spectral sub-sample is characterized by its $(R-I)_c$ color (plotted along the abscissa in the above plot). For the stars in each spectral sub-sample, an RAC is constructed by means of a log-log plot of a chromospheric parameter (CaII emission flux) *versus* a rotational parameter (P/sini): then an RAC slope is determined by least-squares fitting. In the above figure, the RAC slope for each spectral sub-sample is plotted along the vertical axis. The lower curves refer to low-activity stars. The upper curves refer to high-activity stars. In both upper and lower curves, solid (dashed) lines refer to heteroscedastic (homoscedastic) least-squares fits to the data. Adapted from Houdebine et al 2017 (H17).

### 1.3. Comparison of RAC slopes for low-activity stars and high-activity stars

In H17, in order to do a meaningful analysis of the CaII fluxes, our attention was mainly focused on deriving the slope of the RAC at each spectral sub-type. In preparation for this study, the overall sample of 272 stars was separated into sub-samples of "low-activity" (with no detectable Hα emission) and "high-activity" (with measurable emission in the Hα line). It was found that this separation of stars based on activity levels leads to a helpful ordering of the data.

In all points plotted in Fig. 1, we plot (as a filled circle) the mean value of the RAC slope for a particular spectral sub-type, as well as the 3σ standard deviation above and below the mean.

In the lower part of Fig. 1, the plotted data refer to the RAC slopes which were obtained for the low-activity stars, i.e. those with spectral types dK and dM. For these stars, the slope of the RAC was found to be *shallow* (-0.5 to -0.9). It is noteworthy that, for the low-activity stars in our sample, the slopes are found to be relatively shallow in all spectral sub-types which are included in our sample, i.e. ranging from dK4 to dM4. That is, for *all* low-activity stars in our sample, the values of the slopes, as plotted in Fig. 1, remain closer to the horizontal axis than for any other sub-sample of our stars.

On the other hand, when we consider the high-activity stars, the slopes of the RAC were found to follow a more complicated pattern. When we consider the earliest sub-types of these stars, i.e. dK4e, dK6e, and dM2e, the slopes were found to be *steep* (-1.4 to -1.9): see the upper left part of Fig. 1. In contrast, when we consider the two latest sub-types of high-activity stars (dM3e, dM4e), the slope was found to shift towards *shallower* values (-0.9 to -1.2): see upper right area of Fig. 1.

Note on statistics: many of the spectra used in H17 were obtained part from archives of the HARPS spectrograph, which was designed with a view to searching for exoplanets. Because of the small amplitude in velocity which would be induced by a planetary companion, selection of target stars was planned in such a way as to exclude as many sources of noise as possible. One such source is activity on the star: the appearance and disappearance of temporary dark spots and bright plages distort the line profiles of the parent star, thereby obscuring the (small) shifts of the line which might be associated with the presence of a planet. As a result, the selection of target stars for the HARPS instrument were biassed *in favor of* stars with low levels of activity, and biassed *against* stars with the highest levels of chromospheric activity. The existence of this bias explains why, at (e.g.) spectral type M2, the H17 sample contained 54 low-activity stars (with spectral type dM2) while the number of high-activity stars (with spectral type dM2e) amounted to only 11. The bias which is present in HARPS data against high levels of activity should not be interpreted to mean that, in our study, we excluded active stars (i.e. those with spectral types dKe and dMe) entirely: on the contrary, the presence of a significant minority of active stars in our sample specifically enables us to contrast the chromospheric properties of stars where acoustic fluxes may dominate from those where magnetic processes may dominate. An anonymous referee has pointed out that an advantage of including both active and inactive stars in our sample is that "differences in chromospheric heating driven by different dynamo mechanisms may grow more obvious while possible uncertainties/variations in the basal flux level as a function of $T_{eff}$ would be mitigated". In the present paper, we will present results for fluxes of CaII emission in *both* active and inactive stars: the conclusions we will draw turn out to be similar for both groups**.**

Despite the relatively small samples of high-activity stars, it is important to note the statistical significance which is attached to the RAC slopes which are plotted in Fig. 1. This significance will allow us to quantify the differences which exist between the RAC slope among active stars and the RAC slopes in inactive stars.

Specifically, referring to Table 1 in H17, the RAC slope for K4e stars (using homoscedastic least-squares fits) is found to have a value of -1.88±0.55, while for the K4 stars, the RAC slope is -0.62±0.11. Thus, at the 3σ level, our data indicate that the shallowest RAC slope for K4e stars (-1.33) is widely separated from the steepest RAC slope for K4 stars (-0.73): the separation between these extremes in slope is clearly in excess of 3σ.

Similarly, the RAC slope for K6e stars (using homoscedastic least-squares fits) is found to have a value of -1.36±0.16, while for the K6 stars, the RAC slope is -0.52±0.11. Thus, at the 3σ level, our data indicate that the shallowest RAC slope for K6e stars (-1.20) is widely separated from the steepest RAC slope for K6 stars (-0.63), i.e. a separation well in excess of 3σ. And if we examine the same parameters using heteroscedastic least-squares fits, we find that at the 3σ level, our data indicate that the shallowest RAC slope for K6e stars (-0.97) is widely separated from the steepest RAC slope for K6 stars (-0.65): the separation between these extremes in slope is again clearly in excess of 3σ.

Similarly, for M2e stars, the RAC slope (using homoscedastic fitting) is found to have a value of -1.82±0.14, while for M2 stars, the RAC slope has a value of -0.89±0.12. Thus, at the 3σ level, the shallowest RAC slope for M2e stars (-1.67) is widely separated from the steepest RAC slope for M2 stars (-1.01): the separation between these extremes in slope is once again clearly in excess of 3σ. A similar conclusion emerges from examination of the heteroscedastic least-squares fitting for M2 stars. Therefore, for the K6 and M2 stars in our sample, the RAC slopes of the high-activity stars are *steeper*, by a statistically significant amount (> 3σ), than the RAC slopes of the low-activity stars.

However, a different statistical behavior emerges at spectral types M3 and M4. In both of these cases, Table 1 in H17 indicates that at the 3σ level, the RAC slopes *overlap.* (This conclusion is true in both homoscedastic or heteroscedastic fitting cases.) E.g., for M3e stars, the RAC slope is found to be -1.17 ± 0.52 while for M3 stars, the slope is found to be -0.93 ± 0.22. That is, in a statistical sense, the RAC slope for active M3e stars can *not* be distinguished from the RAC slope for inactive M3 stars. Likewise, for M4e and M4 stars, the RAC slopes are found to be -0.96 ± 0.30 and -0.91 ± 0.39 respectively: therefore, again in a statistical sense, the RAC slope for active M4e stars is indistinguishable from the RAC slope for inactive M4 stars.

### 1.4. Interpretation of RAC slopes for stars with low- and high-activity

If there exists an interface between radiative core and convective envelope in any particular lower main sequence, magnetic activity may be powered by an αΩ-dynamo provided that rotational conditions at the interface are favorable (e.g. Mullan et al 2015). If such a dynamo is in operation in a star, the magnetic activity would be expected to show some indications of being sensitive to rotation, i.e. at least some physical quantity which is associated with the "strength" of the dynamo should increase relatively steeply as Ω increases. In an RAC plot, such behavior would be appear in the form of a relatively steep slope. The physical quantity which we favor in the present paper to be representative of the dynamo "strength" in lower main sequence stars is the flux F(CaII) of chromospheric emission in the CaII resonance lines. This leads to an expectation that the RAC for a sample of such stars should have a relatively steep slope. H17 suggested that this could explain the maximally steep RAC slopes which they obtained in dK4e, dK6e and dM2e stars. We note the significant point that stars with spectral types K4, K6 and M2 are believed (see, e.g. Mullan et al 2015) to have masses which are indeed large enough that a radiative core persists in the central regions of the star (Stromgren 1952; Osterbrock 1953), i.e. an interface dynamo is possible in such stars.

On the other hand, an $\alpha^2$ dynamo is expected to be at least in principle capable of operation in *all* low-mass stars because the convective envelope in such stars occupies at least 50% (and in some cases 100%) of the volume of the star. H17 suggested that in the inactive dK4-dM2 stars, the fact that we observe the RAC slopes to be maximally shallow, and significantly (at greater than the 3σ level) shallower than the RAC slopes for the dK6e-dM2e stars, may be attributed to the operation of an $\alpha^2$ dynamo (with its reduced sensitivity to rotation: Durney et al [1993]) in dK4-dM2 stars.

However, in stars where the transition to complete convection (TTCC) has occurred, i.e. the convective "envelope" expands to occupy the entire star, the interface is non-existent, and the αΩ-dynamo is not accessible at all. In such stars, an $\alpha^2$ dynamo may provide the only option for dynamo activity. If this is the case, then the RAC slope for all stars should approach the shallow value which is characteristic of $\alpha^2$-dynamo operation. This expectation is seen in the Figure above as regards M3 and M4 stars.

Based on this interpretation of the *slopes* of the RAC which are plotted in Fig. 1, H17 concluded the following: the TTCC occurs at spectral sub-types between M2 and M3.

**1.**5. Can more than one dynamo operate simultaneously in a given star?

We have already mentioned that dynamos of various types operate in stars: αΩ, $\alpha^2$, and $\alpha^2\Omega$. Is it possible that more than one of these may be operative in a star at an given time? To address this, we note that axisymmetric mean-field dynamo models are described by two coupled equations for the field B and for the vector potential A: e.g. see eqs. 24 and 25 of Charbonneau (2014). Source terms in both of these equations are ultimately responsible for modeling the driving of the dynamo. In the equation for A, only one source term exists: it includes the kinetic helicity parameter α. Because only one source term exists for A, the parameter α plays a crucial role in dynamo action: hence the appearance of at least one power of α in each of the three possible dynamo types. In the equation for B, two source terms $S_1$ and $S_2$ are available: $S_1$ requires the presence of rotational shear (grad Ω), while $S_2$ relies on the presence of a) non-zero value of α. Depending on the parameters in any particular star, either $S_1$ or $S_2$ can be omitted (but not both), and dynamo action may still occur. If $S_2$ is omitted, but $S_1$ is retained, the dynamo is labeled an αΩ dynamo. If $S_1$ is omitted, but $S_2$ is retained, the result is labelled an $\alpha^2$ dynamo. If both $S_1$ and $S_2$ are included, the result is labeled an $\alpha^2\Omega$ dynamo.

In principle, given the structure of the dynamo equations, there seems to be no mathematical reason to state that it would be impossible for two or three of these dynamos to be operating simultaneously in any given star.

Is there any empirical evidence that more than one dynamo is actually at work in a star? As far as we know, no such possibility has yet been reliably reported for the stars of interest to us in the present paper, i.e. K and M dwarfs. However, in the case of the Sun, the possibility of a double dynamo has been raised. Benevolenskaya (1995, 1998) analyzed solar magnetograph data from two complete solar cycles and reported evidence for two main periodic components: one at low frequencies, with a 22 year period, and a second at high frequencies, with a "quasi-biennial" period (i.e. about 2 years). (Analysis of an extensive data set, spanning 160 years, of geomagnetic effects associated with solar activity, indicates that the high frequency component may range in period from 1.2-1.8 years: Mursula et al 2003). Benevolenskaya (1998) suggested that, based on an idea of Parker (1979), two spatially separated dynamos may be operating in the Sun: the 22-yr component near the base of the convection zone (where the radial gradient of Ω is large), while the 2-yr component near the surface (where the latitudinal gradient of Ω is large). Theoretical support for this possibility has been reported by Mason et al (2002) and by Brandenburg (2005).

In view of these results, we should not be surprised if, in any particular K or M dwarf, two different dynamos might be found to be operating simultaneously. If the work of Benevolenskaya (1998) is any indication, the best way to detect two dynamos in a star might be to discover two well-defined periods in the activity cycle of that star. How much difference might exist between such double periods? If the Sun is any indication, the periods might differ by a factor of order 10. Is such a factor detectable in stellar data? To address this, we note that the largest survey of activity cycle periods in low-mass stars which is currently available (for more than 3000 stars) relies on Kepler photometric data which vary on rotational scales (days) as well as on activity cycle scales (years) (Reinhold et al 2017). In the Kepler sample, most of the stars are found to have activity cycles in the range 2-4 years. If a second period is present at 0.1 times the activity cycle, this period could be as short as 70 days: identifying such a short period against the background of rotational modulation (with periods in the range 10-40 days: see Reinhold et al 2017) could be challenging. Nevertheless, it will be interesting to see if, as more data accumulate on activity cycles, any evidence emerges for double periods in the photometric data.

**1.6. The GAIA gap**

Jao et al (2018) have reported, in a sample of some 700,000 GAIA stars within 100 pc, that there exists a gap, i.e. "a small slice of the HR diagram", which is less populated than surrounding regions in the HRD. The slice lies at an absolute K magnitude of -6.7 with a width of only 0.05 magnitude. The slice overlaps in color with single stars having spectral types of M2.0V, M3.0V, and M4.0V. Jao et al noted that the gap lies near the regime where M dwarfs "transition from partially to fully convective, i.e. near spectral type M3.0V". Although Jao et al do not refer to H17, we consider it remarkable that the spectral range M2-M3, which H17 identified as the range where chromospheric heating undergoes a significant change in properties overlaps with the range of spectral types associated with the GAIA gap. This overlap encourages us to undertake a more extensive study of the CaII emission properties than was possible at the time the H17 paper was written: the goal, as in H17, is to determine if there exists an observational signature in CaII data associated with what Jao et al describe as the transition from partly to fully convective, and H17 refer to as the TTCC.

Further reasons to explore empirical data sets in the vicinity of the TTCC for tell-tale signatures of the TTCC is provided by theoretical modelling of M dwarfs. Using a fine grid of stellar model, Van Saders and Pinsonneault (2012: VP12) discovered that, "near" the TTCC, an instability driven by $He^3$ can occur if the (deep) outer convective envelope comes into contact with a (small) convective core. Results for models with masses of 0.35-0.38 $M_\odot$ indicated oscillatory behavior in radii and luminosity. However, the information provided in the VP12 paper does not allow us readily to make comparisons with luminosities or spectral types of the relevant stars.

On the other hand, MacDonald and Gizis (2018: MG18) have presented models which demonstrate clearly a dip in the luminosity function at $M_K$ = -6.7, with a width of close to 0.05 magnitude, thereby replicating very well two of the empirical features of the gap discovered by Jao et al (2018). MG18 find that the narrowness of the gap is associated with the narrow range of stellar masses (between 0.315 $M_\odot$ and 0.345 $M_\odot$) over which there can be a merger between convection zones in the core and in the envelope. Their (implicit) computing method prevents them from identifying any oscillatory behavior of the type reported by VP12. MacDonald (2019) has informed us that, in the range of masses where the merger occurs, his models have $T_{eff}$ ranging from 3450 K to 3480 K: referring to Luhman et al (2003), MacDonald reports that this corresponds to a spectral type near M2.5.

This is a significant result in the context of H17 where we were interested in the possibility of dynamo activity in our sample stars especially in the spectral range M2 to M3. In H17, however, we simplistically assumed a specific possibility, i.e. we thought that an αΩ dynamo could operate effectively in a star lying above the TTCC, but would cease operation in a star lying below the TTCC. Now, however, in light of the VP12 and MG18 modelling, we are faced with a more complicated transition. The existence of an αΩ dynamo has traditionally been associated with the interface (tachocline) between the *bottom* of the outer convection zone and the *top* of the inner radiative region (Spiegel and Zahn 1992). But the models of VP12 and MG18 indicate that we now need to confront the fact that there can be a *second* interface lying much deeper inside stars with low enough mass: this second interface occurs between the *top* of the (small) inner convective zone (core) and the *bottom* of the interior radiative region. According to MG18, for stars with masses between 0.31 and 0.34 $M_\odot$, such a second interface is predicted to exist for time scales ranging from less than 1 b.y. for the 0.31 $M_\odot$ star, to as much as 9 b.y. for the 0.34 $M_\odot$ star. Thus, the existence of such an inner interface is guaranteed for a significant fraction of the age of the universe for M dwarfs in the appropriate mass range. Could a second tachocline exist at the second (inner) interface? And if dynamo activity were to occur in such a tachocline, would the fields thereby generated ever be able to rise to the surface, eventually contributing to chromospheric heating? We have already argued (Mullan et al 2015) that magnetic fields generated at a deep tachocline in a K or M dwarf can indeed by buoyed up to the surface even if the tachocline is located quite close to the center of the star. Therefore, it is at least possible that the existence of a second tachocline in M dwarfs could contribute to magnetic fields at the surface of the star: the presence of such fields, lasting for a time of billions of years, could complicate the response of the chromosphere to magnetic heating. Instead of a sharp transition at a single spectral type, we may need to confront the possibility that the transition might be somewhat smeared out, depending on the age composition of the sample of stars we are examining. In particular, since more massive stars (i.e. 0.34 $M_\odot$) do not become fully convective for about 9 b.y., whereas the less massive stars (0.31 $M_\odot$) lose their "inner tachocline" in the course of a much shorter time (about 1 b.y.), we expect that the *more massive* stars will retain their extra interface "inner" dynamo for a longer time. This would skew the effects of chromospheric heating to last longer in somewhat more massive stars, i.e. those with *earlier* spectral types. To the extent that this is valid, our attempt to identify a signature of TTCC using chromospheric data could be shifted to somewhat earlier types than the spectral type at which the transition to full convection based on a global parameter such as luminosity occurs. Thus, while Jao et al (2018) indicated that the GAIA gap (based on luminosity) is centered in stars with spectral type M3.0, our use of a chromospheric signature (based on magnetic effects at the inner tachocline) might be centered at an earlier type, i.e. closer to M2.0.

But in any case, if the GAIA gap is indeed associated with the presence of double convective regions inside an M dwarf (as interpreted by MG18), then the transition in chromospheric heating between the presence/absence of an αΩ dynamo should occur at a spectral type which is close to that defined by Jao et al, i.e. in the range M2.0-M4.0.

**1.7. The absolute level of CaII emission as a function of spectral type: old data**

In the present paper, we are especially interested in the amount of mechanical energy which heats the chromosphere in our target stars. As well as being interested in the slope of the RAC, we are also interested in the magnitude of the CaII emission flux which is present in each of our spectral sub-samples of K and M dwarfs. In Figure 2 (extracted from a figure in H17), we present an overview of the behavior of the absolute fluxes of CaII emission in stars of increasingly late spectral type. However, since we are interested here in comparing the data with theoretical models of acoustic flux, we restrict the data in Figure 2 to include only the *low*-activity stars (with spectral types dK and dM). As H17 have pointed out, the data for these stars indicate that there is a systematic decline in CaII emission flux as we proceed from mid-dK to dM4. For our low-activity stars, the results in Fig. 2 demonstrate that the decline in going from mid-K to M4 at a given rotational period amounts to a factor of 10 or more. Over the narrower range between M2 and M3, the CaII emission flux declines (at a given period) by a factor of about 2.

In terms of the absolute level of CaII emission in the stars in the H17 sample, we can use the results in Fig. 2 to prepare for a subsequent (see Section 3 below) comparison with theoretical estimates of acoustic fluxes. In this regard, we note that the slowest rotators among the (inactive) dM2 and dM3 stars in Fig. 2 have empirical fluxes F(CaII) (in units of ergs/cm$^2$/s) which are in the range from log F(CaII) ≈ 4.0 to log F(CaII) ≈ 4.3. By definition, the stars in Fig. 2 (all of which are assigned to the category of *low*-activity stars) exhibit no detectable chromospheric *emission* in the Hα line. As a result, the most prominent emission features in the visible spectra of the dM2 and dM3 stars in Fig. 2 are the CaII H and K lines. In view of this, it might at first sight be suspected that the values of log F(CaII) = 4.0-4.3 could be regarded as a zeroth order estimate of the mechanical energy fluxes which are heating the chromospheres in the inactive M2 and M3 stars.

Unfortunately, this conclusion is subject to serious limitations: the absence of Hα *in emission* in M dwarfs is *not* an unambiguous indication that chromospheric heating is absent. Because Hα is not a resonance line, but requires population to be built up in an excited state, it is known theoretically (Cram and Mullan 1979; Cram and Giampapa 1987; Houdebine and Stempels 1997) and observationally (Stauffer and Hartmann 1986; Houdebine and Stempels 1997) that, in the atmosphere of an M dwarf, when the flux of mechanical energy is not too large, the resulting chromospheric heating can enhance the *absorption profile* of Hα without driving the line profile into emission. Without access to detailed modeling, there is no simple method to quantify the mechanical energy fluxes which are needed to explain the empirical amounts of Hα *absorption* in any given inactive M dwarf. To be sure, in an active dMe star, the mechanical fluxes are certainly larger than in an inactive dM star. One example of the requisite quantitative modelling in two dM1 stars, one active, the other less active, has shown (Houdebine 2010) that the energy required to explain the Hα profile exceeds the energy required to account for the CaII H and K lines (plus the CaII infra-red triplet) by a factor of several.

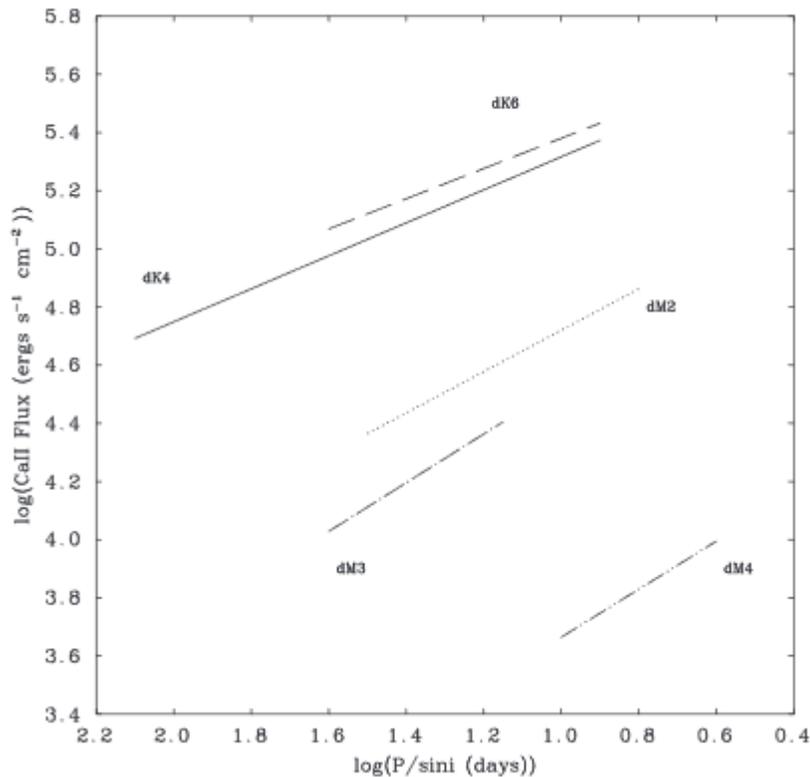

Figure 2. From H17: the flux of CaII emission as a function of rotational period for a sample which contains only low-activity stars (dK, dM).

Furthermore, there are other spectral lines in inactive stars where chromospheric emission is present, e.g. Ly-α and MgII. As regards Ly-α, Doyle et al (1994) have argued that if a basal flux exists in this line in inactive M dwarfs, it is no larger than log F(Ly-α) = 2.9: we can safely neglect this compared to the values of log F(CaII). As regards MgII emission, Mathioudakis and Doyle (1992) have reported that the basal fluxes in MgII in M dwarfs with spectral types of interest to us here lies in the range log F(MgII) = 4.0-4.3: this is essentially identical to the F(CaII) fluxes in the most inactive stars in Fig. 2. To account for radiative losses from other lines of MgII and CaII, Rammacher and Ulmschneider (2003) used fully-time dependent NLTE calculations of the solar chromosphere to show that all line emissions from MgII exceed those in the MgII k line by a factor of 1.4, and all line emission from CaII exceed those in the CaII K line by a factor of 4.3. Houdebine (2010) applied semi-empirical modelling to a dM1e star and found that when all emission lines are combined, the total requisite mechanical energy flux exceeds the CaII fluxes by factors of 10 or more.

Moreover, it is not merely the obviously chromospheric emission lines which must be assessed for mechanical energy deposition: many photospheric absorption lines in the spectrum of an active M dwarf can be partially filled in as a result of chromospheric heating (Houdebine 2010). When model atmospheres are used to quantify the mechanical energy required to fill in the many photospheric absorption lines, Houdebine (2010) found that the total mechanical energy flux exceeds that observed in the CaII H and K lines (plus the energy required for the infra-red CaII triplet) by factors of x = 10-100. Reliable values of x are available only for two more or less active dM1 stars. In the case of inactive M dwarfs, quantitative information is not yet available about the filling-in of photospheric lines by chromospheric heating: however, in such stars, the value of x is almost certainly smaller than the above factors. Thus, in the inactive M dwarfs, the x factor may be less than 10.

As a result, although it would certainly be advantageous to have access to a simple formula which states (e.g.) that the total mechanical energy flux F(mech) = x times the flux in CaII H and K emission, such a formula is difficult to rely on in practice: the factor x can apparently be as large as 10-100 in the most active M dwarfs (Houdebine 2010), or probably less than 10 in the inactive M dwarfs. In view of this, we consider it very difficult to convert our measured CaII H and K fluxes to an *absolute value of* F(mech) in any particular star. If such absolute fluxes *could* be obtained, it would be interesting to compare them with theoretical fluxes of acoustic waves. In Section 3 below, we shall describe the theoretically predicted acoustic fluxes as a function of $T_{eff}$. However, our inability to make comparisons of *absolute values of F(mech)* need not preclude us from undertaking a *differential study* of the following kind: how much do the CaII H and K fluxes *vary* as we move from one spectral type of M dwarf to a closely neighboring spectral type? After all, the observational signatures of mechanical energy deposition in emission and absorption lines in any particular star scale essentially with the *temperature gradient* in the chromospheric structure of that star (Houdebine 2010). In view of this underlying scaling, if we can identify *changes in* the CaII line fluxes from one spectral type to another, these changes should provide (at least roughly) some estimates of the *changes in the* total F(mech) between one spectral type and another. In a *differential study*, there is no need to make individual measurements on *all* of the lines that are affected by the mechanical heating of the chromosphere.

But before undertaking such a differential study, we first turn to the new data which we have now obtained concerning F(CaII) emission in a larger sample of M dwarfs: we shall examine the new data to see if they can strengthen the dynamo conclusions we have drawn from the RAC slopes.

## 2. Expansion of our sample to a larger number of stars

In order to go beyond the results presented in H17, one of the authors (ERH) has recently expanded the sample of CaII surface flux measurements to (roughly) 600 M1-M8 dwarfs, including (about) 130 dM2 stars. The latter number is a factor 2 larger than the sample of M2 stars which was used in H17. In Figures 3, 4, and 5 we present the Ca II fluxes (in units of $10^5$ ergs cm$^{-2}$ s$^{-1}$ Å$^{-1}$) as a function of $T_{eff}$ for the M stars in our expanded sample. We present the data in three separate plots in order to highlight certain properties of the data. In Figures 3 and 4, the plotted points refer only to *inactive* stars i.e. those with spectral types dMx. In the figures, x is a number with two significant digits, i.e. we identify each star by its spectral sub-subtypes. The difference between Figures 3 and 4 has to do with the bin sizes in spectral sub-subtype which are used for plotting. The bins in Fig. 3, between M1.0 and M4.3, are chosen to include stars with only a single sub-subtype: at later spectral types, in order to have enough stars in the sample, the bins in Fig. 3 are widened to include more than 1 sub-subtype. In Fig. 4, all of the bins are wider than in Fig. 3: between M1.0 and M4.5, bins in Fig. 4 are chosen to include stars spanning a range of 4 sub-subtypes: at later spectral types, the bin sizes are expanded to include more than 4 sub-subtypes. In Figure 5, the plotted points refer to *active* M stars, i.e. those with spectral types dMex: in this figure, between M1.0 and M 6.0, we use the same (larger) bins as those in Fig. 4. Note that the vertical scale on Figure 5 is 10 times larger than the scales on Fig. 3 and Fig. 4: this indicates that active dMe stars emit almost an order of magnitude more flux in CaII H and K than the inactive dM stars do.

A note on error bars. In Figure 3, each error bar has a length equal to a certain standard deviation $\sigma$. It is important to note that this $\sigma$ is *not* a measure of the error in the empirical measurement of an individual CaII line flux: such fluxes are actually quite well defined, and the measurements can be made with errors of typically no more than 10-20%. Instead, the $\sigma$ in Fig. 3 arises from the fact that our CaII fluxes represent averages over multiple flux measurements (typically dozens: see Section 1 above). Each dot plotted in Fig. 3 indicates the mean CaII flux of all available observations of dM stars with a particular spectral sub-subtype, while the error bars associated with each dot indicate the standard deviation of the scatter of all measurements at that sub-subtype about the mean. In Figures 4 and 5, the error bars represent $3\sigma$, where the value of $\sigma$ is once again a measure of the scatter of multiple measurements about the mean.

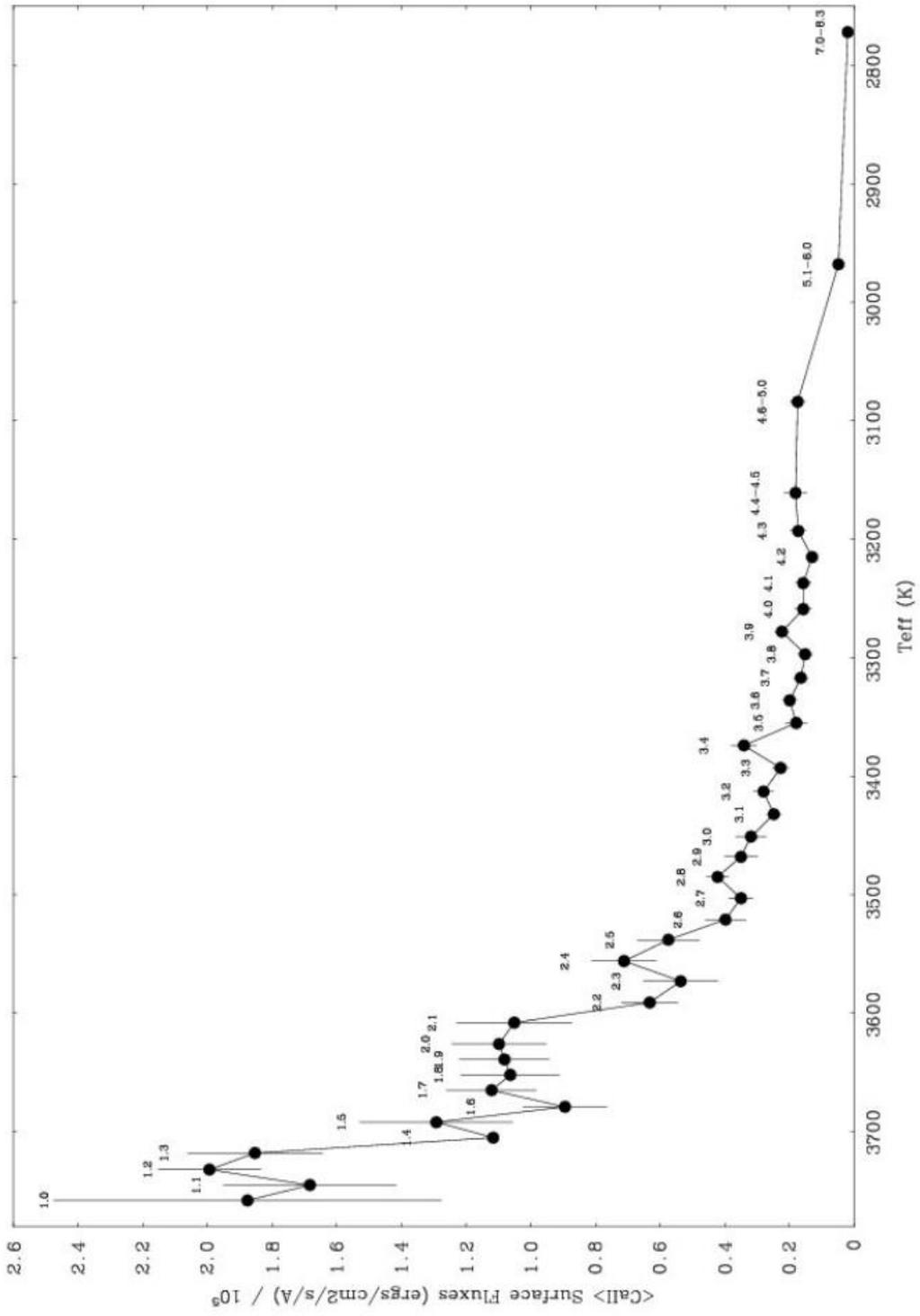

Figure 3. New data: measurements of CaII surface fluxes versus effective temperature for (inactive) M1-M8 stars. The data are plotted in narrow bins of spectral type, in steps of 0.1 times the subtype. Each plotted point is labeled with the spectral sub-subtype of the stars which contribute to that data point.

In plotting the data, we have used as abscissa in Fig. 3 the values of effective temperatures of stars **which were at first estimated** from $(R-I)_C$ colors. **The measurements of stellar colors are now accurate enough that the numerical values of $(R-I)_c$ are reliable** to 3 significant digits. **Using these colors,** Houdebine et al (2019) have reported on a transformation from $(R-I)_C$ to T(eff) which, **when averaged over our sample of hundreds of M dwarfs,** leads to a formal 3σ error of ±39.4 K in T(eff) for M dwarfs. In the present paper, where we quote values of $T_{eff}$ to only 3 significant digits, **we round up the formal estimate to 3σ = ±40 K.**

It is important to note that in deriving values of $T_{eff}$ for each of the stars in our sample, the $(R-I)_c$ colors were used only as a first step. Our final estimates of $T_{eff}$ (Houdebine et al 2019) were obtained by taking the means of the $T_{eff}$ values which we derived from $(R-I)_c$ plus mean values of $T_{eff}$ which have been independently reported in the literature. **The quoted uncertainty (3σ = ±40 K)** is the mean of the difference between these two independently determined temperatures. However, recognizing the difficulties associated with the transformation from color to temperature, and with a view to being appropriately conservative, we can speculate about the possibility of a twofold larger value for the uncertainty: 3σ ≈ 80 K. This would yield σ ≈ 27 K. **We may ask: is a photometry-based value of σ = 27 K in $T_{eff}$ plausible for M dwarfs? To answer this, we note that Kuznetsov et al (2019) have recently reported on a sample of 420 M stars for some of which they list values of $T_{eff}$ based on photometry. For *individual* stars, they list σ values which range from about 40 to 150 K. The average value of σ for an individual M star in the Kuznetsov et al sample is of order σ(1) ≈ 90 K. In our analysis, rather than dealing with individual stars, we work with multiple stars in our sample at every sub-subtype: the number of stars which we have in each spectral sub-subtype can be estimated roughly from Fig. 4, where we give the number of stars in each of our groupings of 4 sub-subtypes. Between M1.8 and M3.3, the average numbers of our sample stars in each sub-subtype is of order n=10. The mean σ(n) value for each sub-subtype is therefore σ(n) = σ(1)/√n ≈ 30 K. Thus, even allowing for a twofold larger value for our uncertainties than the formal value, our estimated uncertainty of 27 K for our samples of multiple stars in each sub-subtype is roughly consistent with the photometry-based results reported by Kuznetsov et al (2019).**

We note that an uncertainty of 27 K in $T_{eff}$ is typically the temperature difference between two adjoining spectral sub-subtypes (i.e. between, say, a star of type M2.2 and a star of type M2.3). When we rely on (R-I) colors in the literature, some of these values are reported in the $(R-I)_K$ system: the formulae which transform the (R-I) colors from the Kron system to the Cousins system yield uncertainties of ≈0.002 mag (Leggett 1992). So when we start with $(R-I)_K$ measurements and transform to $(R-I)_c$, no significant error is incurred. A source of more significant uncertainty is the presence of temporal variations of the R-I color due to activity. However, multiple measurements of R-I are available for many of our stars, and averaging allows us to reduce the activity-related variations. In particular, in low activity stars (i.e. most of the stars in our sample), the variations in (R-I) reported by different observers at different times are typically of order 0.01 mag. Therefore, when we compute averages of the color, temporal variability in R-I (although real) is not a significant source of uncertainty in the low-activity stars which dominate our sample. The principal source of uncertainty in our analysis arises from the calibration of the relationship between R-I and $T_{eff}$ : in this relationship, a large contribution to scatter is due to variations in the metal abundances [M/H] from star to star. In order to minimize these effects, we have excluded sub-dwarfs from our sample.

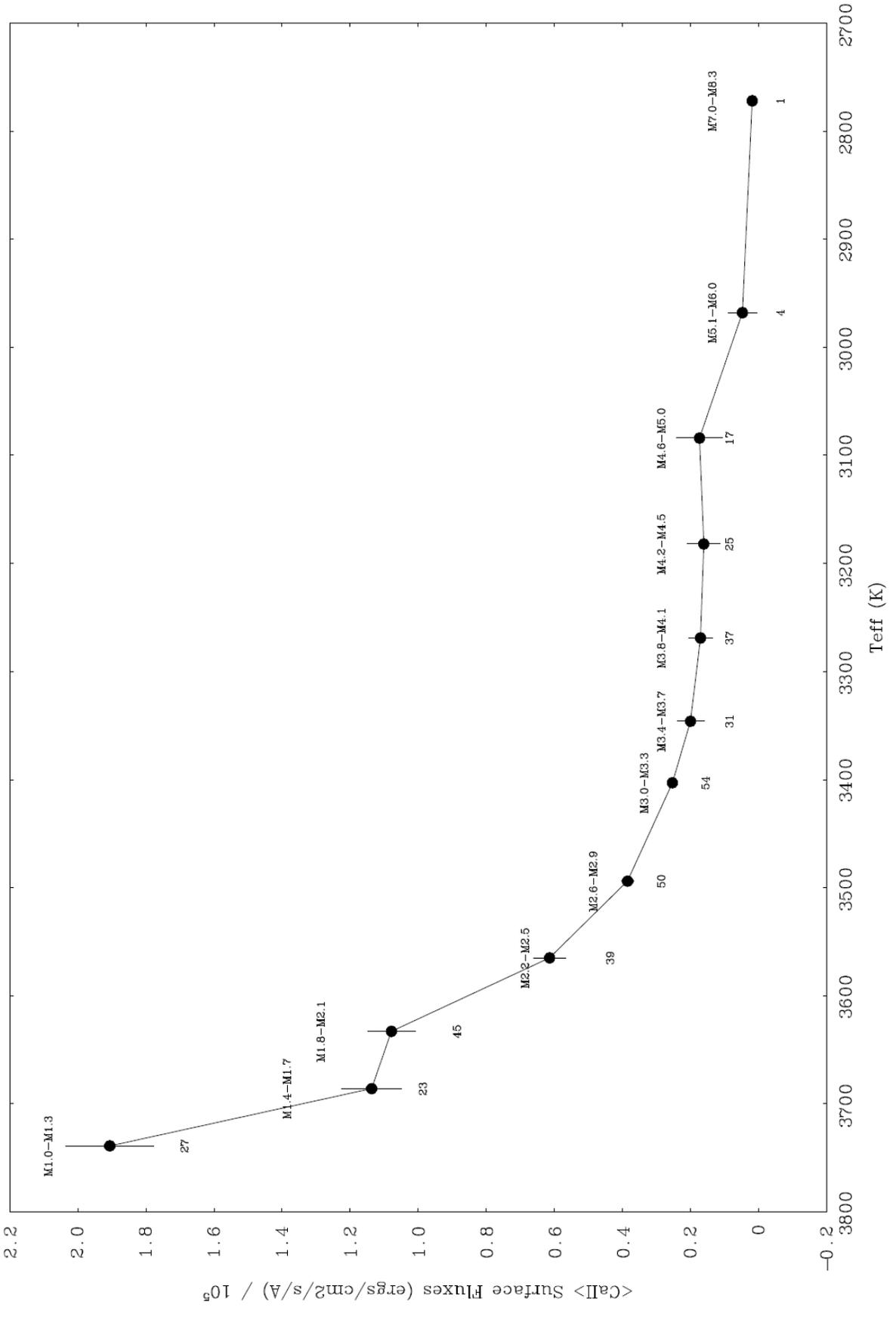

Fig 4. New data: measurements of CaII surface fluxes versus effective temperature for (inactive) M1-M8 stars. The data are plotted in larger bins of spectral type, spanning 4 sub-subtypes. Each plotted point is labeled with the range of spectral sub-subtype of the stars which contribute to that data point. For each point, the number which appears underneath the lower error bar of each point indicates the number of stars in our sample which were used to calculate the data for that point. The largest sample occurs in the bin M3.0-M3.3. The scale along the vertical axis in this figure is the same as that in Fig. 3.

For the stars of most interest to us here, i.e. those with spectral types M1-M3, the $(R-I)_C$ colors reported by Houdebine et al (2019; see their Fig. 3) span a range from roughly 1.0 to 1.5. The corresponding range of T(eff) values is roughly 3750 K to 3400 K, i.e. a range of 350 K. If we wished to define T(eff) values with a **formal** statistical significance of 3σ (=**40** K), we could accommodate about 9 sub-divisions of spectral type into the range of 350 K. In grouping the stars in Fig. 3, we have in fact selected 20 spectral sub-divisions between M2.0 and M3.9, i.e. we have divided the stars into bins each of which spans one-tenth of a spectral subtype (for convenience, we define this as one spectral sub-subtype). For the center of each bin, we have selected the $T_{eff}$ value for each sub-subtype. Thus, the statistical significance of the $T_{eff}$ values in each of our sub-subtypes is not as large as the formal 3σ error mentioned above: instead, each bin corresponds to one-half of the formal error, i.e. about 1.5σ. Although we do not label any of the dots in Fig. 3 to indicate our sample size within a single bin, the sample sizes within the 4-fold wider spectral bins as plotted in Fig. 4 are as follows: M1.0-M1.3 (sample size = 27), M1.4-M1.7 (23), M1.8-M2.1 (45), M2.2-M2.5 (39), M2.6-M2.9 (50), M3.0-M3.3 (54).

A feature in Fig. 3 to which we will draw attention in what follows **(see Section 2.2)** is the relatively sharp decrease in flux between M2.1 and M2.2. The presence of such a sharp decrease suggests that we can be rather confident in the $T_{eff}$ values derived by Houdebine et al (2019) where the estimated uncertainty in $T_{eff}$ indicated that we could assign spectral types with a confidence of 1-2 sub-subtypes. If our assignments had been subject to uncertainties that are larger than 1-2 sub-subtypes, then the sharp decrease in Fig. 3 between M2.1 and M2.2 would have been "smoothed out" due to the incorrect random assignment of stars over several sub-subtype bins. **An anonymous referee has questioned the argument in the present paragraph in the sense that it may involve "a form of circular reasoning and/or confirmation bias": however, in Section 2.2 below, we shall present in Figure 4 the same data which appear in Fig. 3 except that in Fig. 4, the data will be plotted in wider bins. It is important to note that some of the features which we highlighted in Fig. 3 carry over into Fig. 4: the referee considers that such a result "provides a more convincing argument" for the validity of our assignment of sub-subtypes.**

**In a similar vein, the bins in Fig. 3 were selected such that the center of each bin is placed at a single value of the spectral sub-subtype. How sensitive might the shape of the plot in Fig. 3 if we were to shift the bin centers? If all errors in our assignment of spectral type are random, then shifts at the level of one sub-subtype out of a bin in one direction could be compensated by more or less equal shifts in the opposite direction. In such a case, the features ("dips") in Fig. 3 would persist. However, if systematic effects are present (although we are not aware of any such), the compensation might not be exact, and this could introduce spurious "dips" and "pile-ups" in the plots. To address this possibility, we show in Fig. 4 what happens to the data in Fig. 3 when the bins are widened by a factor of 4. It seems quite unlikely that any errors in spectral type assignment could be as large as 4 sub-**

**subtypes: therefore, the plot in Fig. 4 is much less likely to suffer from errors due to a star mistakenly being plotted in the "wrong" bin. In this regard, it is important to note that we shall find (in Section 2.2 below) that at least one of the dips in Fig. 3 (namely, the one between M2.1 and M2.3) persists in Fig. 4. This helps to improve our confidence in the reality of that particular dip.**

We now wish to draw attention to certain aspects of the data in Fig. 3.

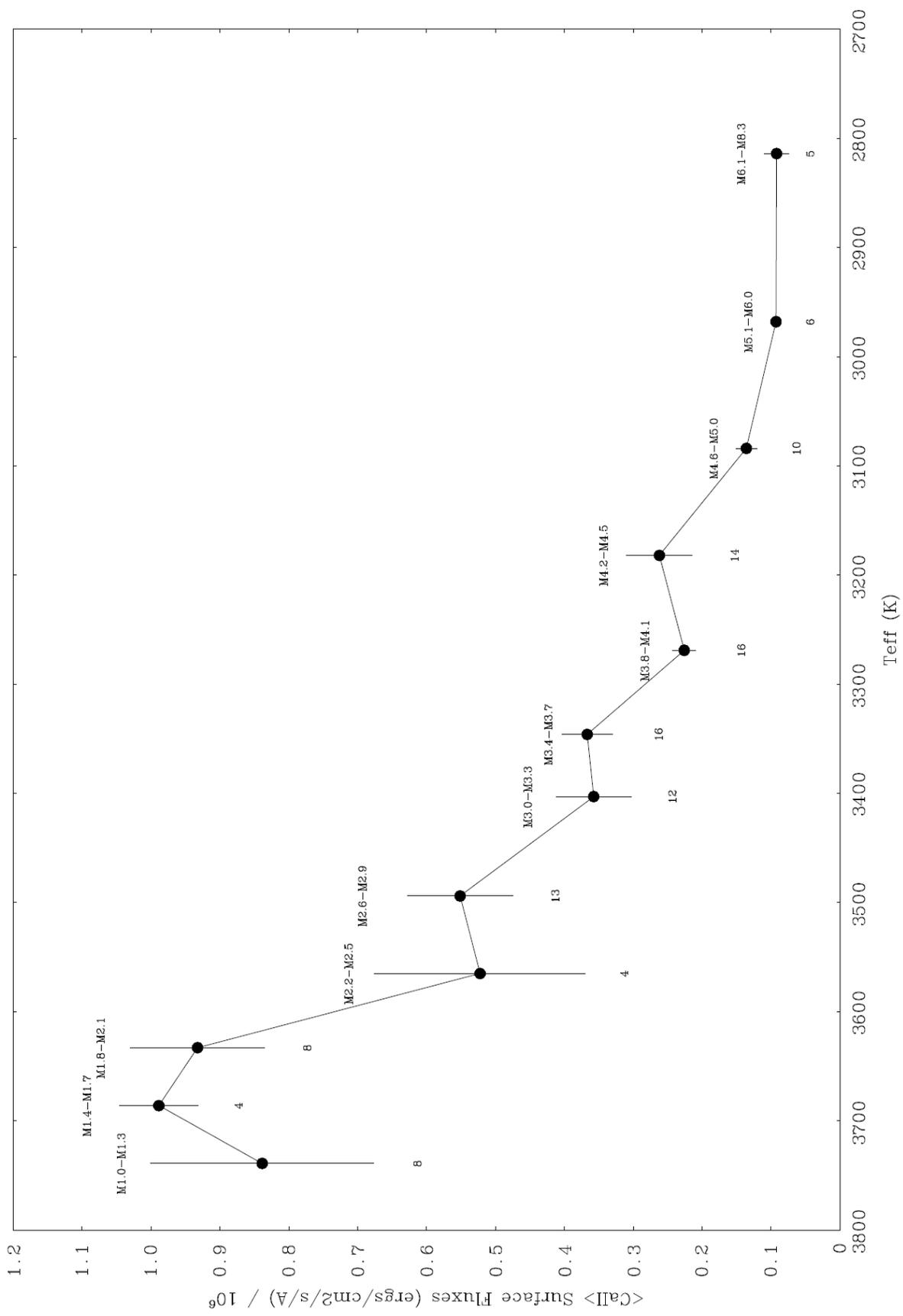

**Fig. 5.** New data: measurements of CaII surface fluxes versus effective temperature for (active) M1e-M8e stars. The data are plotted in the same (larger) bins of spectral type as in Fig. 4. The flux units on the vertical axis are 10 times *larger* than those in Figs. 3 and 4.

### 2.1. The overall decline in Ca emission flux towards later spectral types

It is obvious that there is an overall decline in F(CaII) as we consider stars of increasingly late spectral types. That is, the flux of mechanical energy which is heating the chromosphere decreases as the spectral type increases. Now the flux of mechanical energy which is available for chromospheric heating is derived ultimately from the bolometric luminosity which emerges from the star in radiative form. The ratio of F(CaII) in any particular star to the bolometric flux F(bol) = $L_{bol}/4\pi R_*^2$ of that star is a measure of the fraction $f_{mech}$ of the stellar output which is converted into mechanical form, to be deposited eventually into the chromosphere. The data in Fig. 3 enable us to address the following questions: does $f_{mech}$ vary with spectral type? Or does $f_{mech}$ remain constant in stars with different spectral types?

To address this, we note that F(bol) can be written as $\sigma_B T_{eff}^4$ where $\sigma_B$ is the Stefan-Boltzmann constant. In order to determine how $f_{mech}$ varies with spectral type, let us consider two sample spectral types, M1.5 and M4.2. For these two types, $T_{eff}$ has values of roughly 3750 K and 3250 K (Houdebine et al 2019). Thus, F(bol) has numerical values of $1.1 \times 10^{10}$ and $6.3 \times 10^9$ ergs cm$^{-2}$ sec$^{-1}$ at spectral types M1.5 and M4.2.

As regards the fluxes F(CaII) the data in Fig. 3 (and assuming a line width of 1 Å) indicate that at M1.5, the mean F(CaII) = $1.3 \times 10^5$ ergs cm$^{-2}$ sec$^{-1}$, while at M4.2, F(CaII) = $0.12 \times 10^5$ in the same units. Thus, at M1.5, $f_{mech} = 1.2 \times 10^{-5}$, while at M4.2, $f_{mech} = 1.9 \times 10^{-6}$. These numbers show that the efficiency of chromospheric heating decreases by a factor of about 6 as we go from M1.5 to M4.2. Our data indicate that the flux of chromospheric emission is *not* simply proportional to the bolometric flux: there is more going on in the process of chromospheric heating than simply siphoning off a fixed fraction of the available energy flux emerging through the surface of the star. Instead, the fraction of stellar energy flux which goes into chromospheric heating *decreases* at later spectral types. Later type stars are apparently less efficient at generating mechanical energy from a given flux of radiant energy. If the total radiative losses F(r) from the chromosphere are of order 10-100 large than F(CaII), then F(r) may have values in the range $1.3 \times 10^{6-7}$ at spectral type M1.5, and F(r) may have values in the range $1.2 \times 10^{5-6}$ at spectral type M4.2. In these cases, $f_{mech}$ may be as large as $1.2 \times 10^{-(3-4)}$ and $1.9 \times 10^{-(4-5)}$ at M1.5 and M4.2 respectively.

For comparison, in the quiet Sun (i.e. where heating is not dominated by magnetic effects), the total radiative losses from the chromosphere amount to $4 \times 10^6$ ergs cm$^{-2}$ sec$^{-1}$ (Noyes and Withbroe 1977). Comparing this to F(bol) = $6.3 \times 10^{10}$ ergs cm$^{-2}$ sec$^{-1}$ for the Sun, we find $f_{mech} = 6 \times 10^{-5}$ in the quiet Sun. Thus, red dwarfs with spectral types M1.5 and M4.2 are more efficient at heating the chromosphere than the quiet Sun by factors which may be as large as about 20 and 3 respectively.

## 2.2. "Dips" in the decline in Ca emission flux towards later spectral types.

In Figure 3, we plot data for the (inactive) dM stars, using steps of 0.1 in spectral sub-type. Examination of the plot suggests that, superposed on the overall decline in CaII line flux towards later spectral types (described in Section 2.1), there may be some "fine structure". We wish to determine if this "fine structure" contains physically meaningful information, perhaps associated with the "switching off" of a dynamo mechanism.

At first glance, we note that, in the section of the curve between M1 and M3, "fine structure" might be present in the form of "dips" in the curve at M1.1, M1.4, and M1.6. However, some of these dips consist of only one data point: as a result, it is difficult to rule out the possibility that these are merely noise in the data. On the other hand, there are also dips in the curve which contain *two* consecutive low points, such as at M2.2, M2.3, and at M2.6, M2.7. It seems less likely that these can be considered as merely noise.

In order to identify dips which may be distinguished from mere noise, we consider an argument which is suggested by an analysis that has been used to enhance the confidence of identifying *bona fide* flares in noisy photometric data (see Paudel et al 2018). In the work described by Paudel et al, temporary increases in light level are observed from time to time against a background which is essentially noise. Any particular temporary increase in light level *might* be due merely to noise, or it *could* be due to a *bona fide* flare. However, if two increases in the light level occur right next to each other, i.e. during two contiguous intervals of observing, the chances of those two increases in photometric level being a *bona fide* flare are significantly improved compared to the case of an isolated (single) increase in the photometric level. Paudel et al use data on the amplitudes $A_i$, $A_{i+1}$ of pairs of successive events and normalize each amplitude to the local standard deviation of the noise levels $\sigma_i$ and $\sigma_{i+1}$. The product ($A_i/\sigma_i$) times ($A_{i+1}/\sigma_{i+1}$) is considered to provide a quantitative measure of the statistical significance of the possibility that a flare has in fact been detected. Let us consider two possible cases. In case (i), where all that is observed is a single "blip", the amplitude of that blip $A_i$ may be large compared to $\sigma_i$, but the succeeding data point will have an amplitude $A_{i+1}$ which is small compared to $\sigma_{i+1}$. As a result, the product ($A_i/\sigma_i$) times ($A_{i+1}/\sigma_{i+1}$) in case (i) will turn out to be relatively small: in such a case, one concludes that a *bona fide* flare did not occur. On the other hand, in case (ii), where two successive amplitudes are observed to be large, then the product ($A_i/\sigma_i$) times ($A_{i+1}/\sigma_{i+1}$) is a relatively larger number than in case (i): in such a case, the chances that a flare really did occur are greatly improved. We do not claim that our approach here is as statistically rigorous as that described by Paudel et al (2018): unlike Paudel et al, we do not have the luxury of dealing with a background which is essentially noise. Moreover, we are not dealing with a *time* series of data: instead, we are examining a series of data arranged in order to spectral subtype. Nevertheless, we consider that our examination of observations grouped in pairs of immediately adjacent spectral subtypes captures the spirit of the approach of Paudel et el (2018).

With this in mind, we start with the data points in Fig. 3, i.e. at each spectral sub-subtype, we have a value of the flux $F_i$ and a value of its associated standard deviation $\sigma_i$. Using these, we calculate the difference between successive pairs of points $\Delta = F_i - F_{i+1}$ and then taken the ratio of $\Delta$ to the combined standard deviation $\sigma = \sqrt{(\sigma_i^2 + \sigma_{i+1}^2)}$. Examining the values of $\Delta/\sigma$ as a function of spectral type for evidence of "dips", we have no interest in pairs of data points for which $\Delta/\sigma$ is a *positive* number. Instead we are most interested in those pairs of data points with the largest (in absolute terms) *negative* numerical value of Δ/σ . We find that the largest negative Δ/σ has a value of -2.9 for pair (a) at spectral sub-subtypes M1.3 and M1.4. The second largest negative Δ/σ has a value of -2.3 for pair (b) at spectral sub-subtypes M2.1 and M2.2. No other pair of successive points has a Δ value as large as 2σ. However, we note that pair (c) at M2.4-M2.5 and pair (d) at M2.5-M2.6 have Δ/σ values of -1.0 and -1.2 respectively. Now we proceed to the second step of our calculation. We consider the product of the above largest values of Δ/σ with the value of Δ/σ for a neighboring pair. Combining pair (a) with the pair M1.2-M1.3, we obtain a product of the two neighboring Δ/σ values equal to +1.8. Combining pair (b) with the pair M2.2-M2.3, the product is found to be +1.3. And combining pair (c) with pair (d), the product is found to be +1.2. These are the only pairs for which the products are in excess of +1.

On the basis of this discussion, we consider that the best candidates for "dips" of interest in Fig. 3 are as follows. The dips which we label as Da, Db, and Dc are found to occur at spectral types M1.2-M1.4, M2.1-M2.3, and M2.4-M2.6.

Now let us consider the data in Fig. 4, where the same sample of (inactive) dM stars which were used for Fig. 3 are grouped into wider bins. The use of wider bins has the effect that the standard deviation of each bin is smaller than those in Fig. 3: the smaller values of σ will lead to (numerically) larger values of Δ/σ. For ease of reference, we will label the points in Fig. 4 in order from left to right as A,B,C... Repeating the exercise described above, we find that the largest negative value of Δ/σ = -5.2 occurs between points C and D. The next largest negative value of Δ/σ = -5.0 occurs between points A and B. The remaining values of Δ/σ become progressively smaller (in magnitude) as we go beyond point D. When we take the next step in our analysis, and calculate the product of Δ/σ for neighboring pairs, we find that the largest product of Δ/σ for neighboring pairs is +21 for the pair C-D and D-E. For the pair A-B and B-C, we find that the product is +3. These results suggest that the best candidate for a "dip" in Fig. 4 extends from C (=M1.8-M2.1) to E (=M2.6-M2.9), with the largest "dip" occurring between C and D, i.e. between M1.8-M2.1 and M2.2-M2.5. Compared to the results in Fig. 3, this range of "interesting" spectral types we have obtained from Fig. 4 overlaps best with "dip" Db identified in Fig. 3, although there is also some overlap with "dip" Dc. Our analysis of Fig. 4 suggests that there is no significant overlap with "dip" Da in Fig. 3.

Turning now to Fig. 5, we examine the data for the (active) dMe stars, which are grouped into the same set of bins as were used in Fig. 4. We will use the same notation as in Fig. 4, labelling each point from left to right in Fig. 4 as A, B, C… We recognize that our sample of dMe stars is smaller than for the dM stars, but we can repeat the analysis that we have applied to Figs. 3 and 4. When we do that, we find that the largest (negative) numerical value for Δ/σ in the spectral range M1-M3 is found to be -2.3, between points C and D. We note that the largest value of Δ/σ that we found in Fig. 3 also occurred between points C and D, just as we find in Fig. 5. Moving on to the next step in the analysis, we find that the product of Δ/σ values for neighboring pairs, we find that the only case where the product is positive occurs for the pair C-D and B-C. Thus, the best candidate for a "dip": in Fig. 5 extends from B (=M1.4-M1.7) to D (M2.2-M2.5). Compared to the results in Fig. 3, we see maximum overlap with dip Db; there is less overlap with dip Dc, but none with dip Da.

The data with the finest resolution (Fig. 3) suggests that the best candidate for a significant "dip" lies between M2.1 and M2.3. This dip is consistent also with our analysis of Figs. 4 and 5. For that reason, we consider that the range M2.1-M2.3 is the best candidate for the location of a change in dynamo mode among early M dwarfs. Whereas H17 suggested that the change in dynamo mode occurred between M2 and M3, we now suggest that we can narrow the range down to M2.1-M2.3.

### 3. Comparison with theoretical work: basal fluxes.

In solar-like stars, and in lower-mass stars, it is well known, based on RAC data, that magnetic activity contributes to enhancements in chromospheric emission in the most active stars. However, an important empirical feature of the chromospheric data is that there exists a *lower limit* on chromospheric activity (as indicated by CaII emission) in cool stars (Schrijver, 1987; Rutten et al. 1991). As far as we know, there are no lower main sequence stars with spectral types of K or M which exhibit *zero* chromospheric emission. The existence of a firm lower limit on chromospheric emission in cool stars led Schrijver and colleagues to the proposal that there exists a "basal flux" of energy which provides a "floor" on the chromospheric heating in all cool stars, even in the least active stars. Now a salient aspect of the sample of stars which have been analyzed in the present work is that, because they are extracted mainly from HARPS data (but also include data from FEROS, SOPHIE, NARVAL, and UVES), the sample of stars is somewhat biased towards low-activity stars. In view of this, we expect that some of the stars in our sample might have chromospheres which lie near the basal level.

In the context of low-activity stars, and because we are interested in pushing observations to the lowest possible limits of flux in the CaII emission line, it might be asked: have any of our target stars been found to have *no* chromospheric emission in CaII at all? This question can be answered definitively: in all of the 600 (or so) M dwarfs for which we have examined archival spectra from HARPS, FEROS, SOPHIE, NARVAL, and UVES, a measurable emission flux has been found in the CaII line. The fact that in all 600 stars, chromospheric emission is measurable in CaII suggests that chromospheric heating relies on a physical process which is present in all of the stars in our sample.

One feature which occurs in all M dwarfs is a deep convective envelope where pressure fluctuations are inevitably associated with convective overturning: these fluctuations generate acoustic waves with a range of periods, and waves of short-enough periods (i.e. shorter than the acoustic cut-off period: see, e.g. Mullan [2009]) can provide a finite amount of mechanical flux to the chromosphere. This flux of short-period acoustic waves sets a firm lower limit on the mechanical energy flux which is available to the chromosphere in any star with a convective envelope. The existence of a lower limit on the acoustic energy flux which reaches the chromosphere is consistent with the concept of a "basal flux" as proposed by Schrijver (1987). A natural question is: is there any justification for the possibility that "low activity" in an M dwarf can be considered as being equivalent to "near the basal level" of acoustic power? To test this, quantitative agreement between acoustic heating and the lowest empirical values of chromospheric emission have been reported by (e.g.) Mullan and Cheng (1993) and by Fawzy and Stepien (2018). To be quantitative, we note that the phrase "near the basal level" is defined as follows in the study by Mullan and Cheng (1993). An empirical value of the basal flux in MgII emission had previously been reported by Rutten et al (1991) as a function of B-V color. For each target star in an available sample of 30 stars, Mullan and Cheng (1993) inserted the B-V color in order to extract a MgII basal flux: then the stars which were to be modeled with acoustic heating were selected to be no more than a certain limit above the basal flux. In order to obtain a sample of stars which was not too small, the limit was set at 0.9 in the log(flux). That is, the stars to be fitted were chosen to have MgII fluxes which exceeded the basal flux by less than one order of magnitude. Then an acoustic model based on dissipation of weak shocks was fitted to the MgII emission fluxes, as well as to the emission fluxes in Ly-α. Satisfactory fits were obtained for both of these individual lines.

However, the work of Mullan and Cheng (1993) was published before it was realized that the total radiative losses from the chromosphere exceed those in individual lines by factors which range from a few up to values as large as 10 or more (Houdebine 2010). In view of the latter development, it no longer seems to be permissible to claim that acoustic power alone is sufficient to provide for the total radiation losses from the basal flux stars. It seems that something in addition to acoustic power may be required to explain the basal flux.

In order to address this with more precision, it is necessary to quantify in more detail the acoustic energy fluxes which can be generated in the atmospheres of low-mass stars.

According to the work of Lighthill (1952) and Proudman (1952), the flux of acoustic power $F_A$ which is generated by turbulence in which the mean speed is v scales as $v^8$. Models of lower main sequence stars (e.g. Castellani et al 1971; Mullan 1971) suggest that the maximum convective velocity v(max) in the models of main sequence stars has a well-defined behavior: there is a peak among A-type stars (log T ≈ 3.9), and then there is a monotonic decrease in v(max) as we go down the main sequence towards cooler values of T(eff).

Based on this, and adopting a model of turbulent eddy distributions in stars along the lower main sequence, a value can be calculated for the flux of acoustic power that is to be expected as we go to lower mass stars on the main sequence. Extensive results have been presented by Ulmschneider et al (1996) for a variety of surface gravities from log g = 0 to log g = 8. A sample of their results is presented in Figure 6: the results in the figure refer to convective zone models which are computed with a value of mixing length = 1.5 times the pressure scale height. In the present paper, we are interested mainly in stars with spectral types M0 to M7, where masses range from about 0.6M(sun) down to about 0.1M(sun). For such stars, the radius and mass are correlated according to R ~ $M^b$ where empirical data indicate that b = 0.945+-0.041 (Demircan & Kahraman 1991). The surface gravity g therefore scales as $M^{-0.89}$. For the M stars which are of most interest to us, this leads to log g values in the range from 4.7 to 5.2.

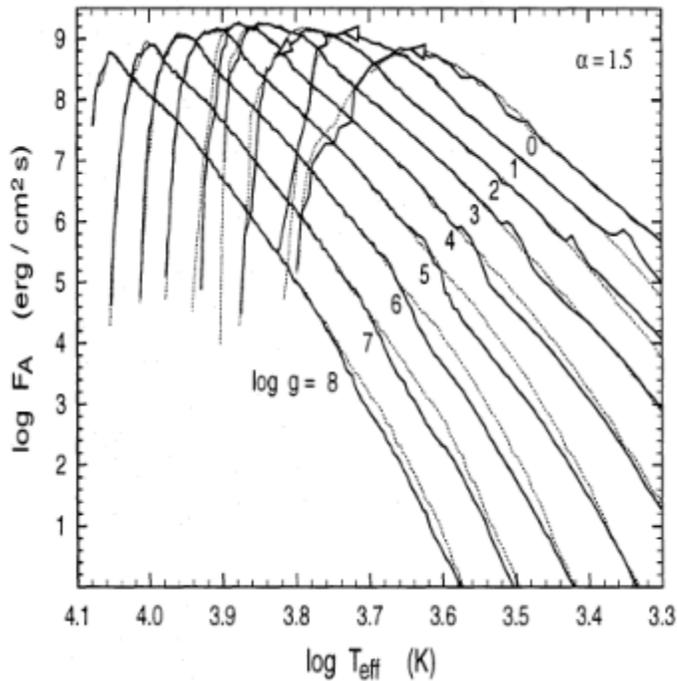

Figure 6. (from Ulmschneider et al 1996) Theoretical evaluation of acoustic fluxes $F_A$ emitted by stars with effective temperatures ranging from 2000 K to 13000 K and gravities in the range from log g = 0 to log g = 8.0. Solid lines: acoustic fluxes from convective models computed with mixing length theory using a mixing length of 1.5 pressure scale heights. Dotted lines: calculated fluxes neglecting the presence of $H_2$ molecules. We are especially interested in results with T(eff) values in the range log(T(eff)) = 3.53-3.57, and gravities of order log g = 5.

The results in Fig. 6 need to be handled with some caution. Although Ulmschneider was the lead author of the paper which generated the results in Fig. 6, he and his colleagues, at a later date (2003), stated that the results "may" lead to an incorrect evaluation of chromospheric heating (Ulmschneider et al 2003: U03). The results in Fig. 6 were obtained by means of a 1-D model in which multiple shocks merge into one another in the atmosphere of a star: such mergers are a natural occurrence if all gas motions are forced to be confined strictly to a single dimension. However, in the light of 3-D considerations, U03 stated that shock merging is expected to occur only rarely. U03 emphasized that "adequate" calculations of chromospheric heating "require a 3-D radiative-hydro code with multiple input sources of acoustic wave spectra", as well as a "fully time-dependent treatment of …ionizations and thermodynamics". U03 stated that such calculations "are not yet feasible with present computer power". In a subsequent paper, Hammer and Ulmschneider (2007) indicate that 1-D models (such as those in Fig. 6) can describe strong brightenings in the chromosphere "quite well", but the "overall chromospheric dynamics" may be governed by 3-D shock propagation.

Because of the complexities of the 3-D problem (as stressed by U03), the results in Fig. 6, spanning as they do 8 orders of magnitude in gravity, and spanning almost the entire range of $T_{eff}$ for main sequence stars with convective envelopes, have never, as far as we are aware, been re-computed using a full 3D code. If such re-computations were to become available over the range of parameters shown in Fig. 6, we would certainly like to compare the results with our observations of CaII emission. In the absence of such results, the results in Fig. 6 cover such a broad range of parameter space that they offer us an opportunity to be used in a narrowly confined *differential study of chromospheric heating,* at least for a zeroth-order approach to the problem. By "narrowly confined" we mean that our data are confined to a narrow range of log $T_{eff}$ (3.5-3.6), and to essentially a single value of log g (5.0).

Using a mean value of log g = 5, we see in Fig. 6 that the values of $F_A$ have a peak at log T ≈ 3.9, and $F_A$ declines towards later spectral types. In the parameter range of interest to us here, i.e. log $T_{eff}$ = 3.6-3.5, and log g = 5, Fig. 6 shows that $F_A$ has values between (about) $10^5$ and $10^3$ ergs cm$^{-2}$ s$^{-1}$. We note that the values of F(CaII) in our inactive sample stars (see Figs. 3 and 4 above), range from about 2 x $10^5$ and 1 x $10^4$ ergs cm$^{-2}$ s$^{-1}$ . Thus, the theoretical acoustic fluxes do overlap with the empirical CaII emission fluxes. However, since the entirety of chromospheric heating requires an energy flux of x times F(CaII), where x may range from a small as a few to as large as 10 or more, then the acoustic fluxes in Fig. 6 *might not* suffice to account for the *mean level* of chromospheric heating. However, in this regard, it is worth noting that the data in Figure 4 represents *mean* fluxes for inactive dM stars: individual stars are scattered about the mean at each spectral type. In fact, M dwarfs with the lowest levels of activity, and also M subdwarfs (which were not used in forming the means in Fig. 4), are observed to have CaII fluxes which lie *below* the mean values plotted in Fig. 4 by factors of 5-10 (Houdebine et al. 2020). Such stars would be the best qualified to be labeled as basal flux stars, and they *could* be powered by acoustic fluxes.

If indeed acoustic power is not sufficient to power the mean levels of chromospheric heating in inactive M dwarfs, we need to ask: where does the rest of the chromospheric heating come from in such stars? From observations of the Sun, it is well known that chromospheric heating is stronger in regions with stronger magnetic fields (Withbroe and Noyes 1977): in going from quiet Sun to an active region, the total radiative losses from the chromosphere increase by a factor of 5. It seems likely that magnetic effects are also at work in heating the mean chromospheres in the inactive M dwarfs. Several magnetic processes have been proposed to heat stellar chromospheres including transverse waves on a magnetic flux tube (Musielak and Ulmschneider 2002) and Alfven waves (Cranmer and van Ballegooijen (2005). Other mechanisms include dissipation of electric currents (Goodman 1995) and nanoflares (Parker 1988; Jess et al 2014). Unfortunately, making predictions based on magnetic effects requires the introduction of many more free parameters than the number which is required for the acoustic fluxes shown in Fig. 6. As a result, the parameter space with 2 dimensions ($T_{eff}$ and g) which enabled Ulmschneider et al (1996) to compute the results in Fig. 6 must be expanded to a larger number of dimensions in order to include magnetic effects. The expanded number of parameters includes at least the following: numerical values of the magnetic field strength, the topology of the field (open? closed? multipolar?), the fractional area of the surface occupied by fields, the efficiency of exciting each magnetic mode on a flux tube, the frequency spectrum of the waves which are excited (e.g. Cranmer and van Ballegooijen consider a spectrum extending from periods of 3 seconds to 3 days, i.e. a range of 5 orders of magnitude), the local geometry of the field lines which carry the wave upwards to the chromosphere (determining the effectiveness of refraction and reflection of the various modes of upward waves [Osterbrock, 1961]), the process which allows each magnetic mode to dissipate in the chromosphere, the processes which enable dissipation of electric currents in a partially ionized magnetized plasma, and, finally, the dimensions in space and time and energy of nanoflares (Jess et al 2014). This large array of unknowns pertaining to magnetic heating of the chromosphere gives rise to a parameter space which is so vast that exploration of even a small volume will require computing enormous effect. Many different combinations of the various parameters might eventually be found which give agreement with some of the data, but It is not clear how one will be able to decide which combination is the correct one for any given star. In contrast, the 2-parameter space represented in Fig. 6 contains information which can be meaningfully used to set at least a lower limit on the magnitude of the mechanical energy that is available for chromospheric heating.

## 4. Possible origin of a "dip" in our CaII line flux data

As regards the decline towards later spectral type in Fig. 6, it is worth noting the prominent "molecular feature" in Figure 6: two sets of results are presented, one plotted with solid lines, the other with dotted lines. The difference between these two sets has to do with the inclusion (solid lines) or non-inclusion (dotted lines) of $H_2$ molecules in the gas. In the case log g = 5, which is the case of primary interest in the present paper, the molecular feature does indeed cause a significant decrease in the acoustic flux over a certain range of T(eff), namely, for log T(eff) between (roughly) 3.65 and (roughly) 3.59. In this range of temperatures, the predicted value of $F_A$ in the molecular inclusion model declines by a factor of about 2 relative to the non-inclusion model. At values of T(eff) lower than 3.59, i.e. after traversing the "molecular feature", the values of $F_A$ for the inclusion model at log g = 5 reverts to a steady decline as log T(eff) decreases, with a slope which is essentially identical to the slope that was in place before traversing the "molecular feature".

With regard to the molecular feature, when we examine the stars which are of most interest to us in this paper, i.e. M1-M3 stars with T(eff) in the range of roughly 3750-3400 K (see Fig. 3), we note that the corresponding values of log T(eff) =3.53-3.57 lie on the cool side of the "molecular feature" in Fig. 6. As a result, we do not anticipate that the "molecular feature" should lead to any significant features in our data. On the contrary, we expect that the predicted values of acoustic flux $F_A$ in our sample of stars should undergo a smooth steady decline as we go from the warm end of our sample (log T(eff) = 3.57) to the cooler end (log T(eff) = 3.53). In fact, over the range of T(eff) values which are of interest to us here, the predicted values of log $F_A$ (with log g = 5 in Fig. 6) can be seen to decline smoothly from about 4.6 to about 3.8, i.e. by a factor of about 5.

In this regard, we consider it a matter of some interest that, according to the results in Fig. 6, as we go from a star with T(eff) = 3750 K to a star with T(eff) = 3400 K at log g = 5, the theory suggests that the decline in $F_A$ should be *smooth*, i.e. more or less monotonic. The predictions of $F_A$ reported by Ulmschneider et al (1996) do not indicate that there should be "dips" or "recoveries" in the decline. Admittedly, a careful examination of Fig. 6 does show that, as we traverse the "molecular feature", there are two regions of the $F_A$ curve for log g = 5 (around log(T(eff) = 3.595 and 3.58) which have shapes that might be classified as plateaus.

In this regard, we wish to draw attention to the empirical features ("dips") which we described in our data in Section 2.2 above. As mentioned there, we consider that the "dip" which has the largest statistical significance is the one between M2.1 and M2.3. What could give rise to such a feature? We recall that the stars in our sample are low-activity stars: such stars are in general believed to be good candidates for acoustic heating of their chromospheres. However, in view of the discussion in the previous section, there is no reason to believe that, over a small range of T(eff) values, the generation of basal fluxes of acoustic power should give rise to dips of acoustic heating for stars which (i) lie within the range of T(eff) that is relevant to our sample, and (ii) have an amplitude as large as that which occurs in our CaII flux data.

This leads us to revert to the possibility that acoustic power is not the sole contribution to chromospheric heating in our sample of stars. At first sight, this might seem surprising, since our sample is biassed towards low activity stars. But the presence of any finite magnetic field (whether generated by an αΩ dynamo or by an $α^2$ dynamo or by an $α^2$Ω dynamo) om a convective envelope will ensure that MHD waves are also generated. In H17, we have argued, based on RAC data, that the steeper RAC's which are observed in the high-activity stars with spectral types dK4e, dK6e, and dM2e stars can be associated with the operation of an αΩ dynamo in these stars. (Of course, the presence of an extensive convective envelope in all M dwarfs means that an $α^2$ dynamo can also be in operation.) The argument for an αΩ dynamo in dK4e-dM2e stars is based on the fact that structural models of such stars (Stromgren 1952; Osterbrock 1953**)** indicates that an interface exists between a radiative core and the convective envelope. But in H17, we also argued that stars of high activity at later spectral types (dM3e, dM4e), do *not* have steep enough RAC's to be considered as sites of αΩ dynamos: instead, the overlap of the RAC slopes for the latest active stars with slopes of the RAC's for inactive stars (dK4-dM4) led us to conclude that an $α^2$ dynamo (or perhaps an $α^2$Ω dynamo: after all, the RAC does have a non-zero dependence on Ω) would be a better candidate to explain the RAC slopes in inactive stars *as well as in the latest (dM3e, dM4e) active stars*. Based on the data in H17, the effects of an αΩ dynamo were hypothesized to "switch off" at a spectral type which lies between M2 and M3. In H17, the "switching off" of the operation of an αΩ dynamo was presumed to be associated with the transition to complete convection (TTCC): however, this transition is now known to be more complicated (Jao et al 2018; MacDonald and Gizis 2018), so the conclusion of H17 needs to be re-stated in a more nuanced manner.

### 4.1. Distinguishing between different processes of chromospheric heating

When there is a possibility that more than one method of chromospheric heating may in principle be capable of operating in a star, the resulting heating of the chromosphere will in general not be expected to give rise to identical amounts of heating. Instead, if N methods are permissible, then method i will give rise to a level F(i) of CaII emission, where i varies from 1 to N. An important quantitative question arises: by how much might F(i) differ quantitatively from F(j)? The range of possible answers can in principle be arbitrarily large. For the sake of simplicity, we consider here only the case N = 3, i.e. acoustic and two different magnetic components. Again for simplicity, we assign one magnetic component to an αΩ dynamo, and the other to an $α^2$ dynamo. Then we assume that the three components of heating lead to emission fluxes in (say) CaII of F(a), F(αΩ), and F($α^2$).

The values of F(a) can be calculated with some confidence for a low-mass star with a specified effective temperature and gravity (see Section 3, where these fluxes were labelled by the notation $F_A$, in accord with the notation used by Ulmschneider et al 1996). However, the calculation of magnetic fluxes cannot be achieved with comparable confidence because of unknown parameters. The only thing that we can say with certainty about the magnetic fluxes ($F(\alpha\Omega)$ or $F(\alpha^2)$) is that both are non-negative. For present purposes, we would like to determine if we can say anything plausible about the ratio of $F(\alpha\Omega)$ to $F(\alpha^2)$. In general, with different processes at work, $F(\alpha\Omega)$ and $F(\alpha^2)$ might differ from each other by significant factors.

In support of this conclusion, we may cite two different studies of dynamo properties. First, Durney et al (1993) have reported on a model of an $\alpha^2$-dynamo, i.e. one in which magnetic field is generated by turbulent motions in the bulk of the solar convection zone (SCZ). In the presence of small-scale turbulence, when there exists an initial weak magnetic field, nonlinear transfer between magnetic field and velocity causes the magnetic energy on short length scales to build up as time progresses. Even in the absence of rotation, the magnetic energy on small scales ($ME_S$) eventually reaches a level almost as large as (within a factor of 2) the equipartition value based on the kinetic energy on small scales ($KE_s$): in the presence of rotation, the approach to equipartition occurs on shorter time-scales. On the other hand, there is also a build-up of magnetic energy on large scales ($ME_L$): however, Durney et al find that the numerical value of $ME_L$ turns out to be considerably smaller than $ME_s$. In the absence of rotation, the value of $ME_L$ rises to only about $0.01 ME_s$. Thus, this model predicts that turbulent convection leads preferentially to magnetic energy on *small* scales. Magnetic fields which are generated at the radiative convective interface depend on a different physical process: in this case, rotation plays an essential role. To describe the process, Durney et al (1990) modeled a dynamo which they consider as closer to the solar cycle case: this dynamo lies near the base of the SCZ, where differential rotation contributes to an $\alpha\Omega$ dynamo. In this case, they found that the magnetic energy can grow to super-equipartition values. They referred to this as a "cycle field", because it helps to account for the 11-year solar cycle. But Durney et al (1990) stressed that "the SCZ could…generate a magnetic field with different properties than the cycle field", i.e. $F(\alpha\Omega)$ could differ from $F(\alpha^2)$. The sense of the difference was not mentioned by Durney et al. Second, Mason et al (2002) have examined the properties of two dynamos in which the α-effect is concentrated (a) at the surface, and (b) near the base of the SCZ. They find that the dynamo is considerably more effective in case (b), i.e. when the α-effect is located close to the interface. Mason et al conclude that an αΩ dynamo operating near the base of the SCZ is "considerably more effective" than a surface $\alpha^2$ dynamo. In this case, it seems likely that $F(\alpha\Omega)$ exceeds $F(\alpha^2)$.

Whatever the individual values of F(a), $F(\alpha\Omega)$, and $F(\alpha^2)$ happen to be, the total amount of chromospheric emission flux which will be detected from any particular star is $F(tot) = F(a) + F(\alpha\Omega) + F(\alpha^2)$. At any particular spectral type, we can firmly assert that F(a) is always present at the level which was quantified in Section 3. We can also assert that, since all the stars we study have deep convective envelopes, $F(\alpha^2)$ will be present to some extent in all of our stars at some level: what that level is in any particular star will probably depend most on how fast the star is rotating. Since this component is always present in late-type stars, it is possible that $F(\alpha^2)$ may contribute to the basal flux (in addition to acoustic power): this possibility is suggested by our findings in Section 3 above that acoustic power (as calculated by Ulmschneider et al 1996) is not sufficient to account for *all* of the radiative losses in basal flux stars. And as regards $F(\alpha\Omega)$, we can say that it will be present in stars which are massive enough to have a radiative core, but it will be absent in completely convective stars.

Some information about the relative magnitude of F(αΩ) and F($\alpha^2$) can be gleaned by considering stars belonging to a category in which both processes are potentially at work. We have already pointed out (see Fig. 1 above), that inactive stars with spectral types dK4, dK6, and dM2 have RAC slopes which are definitely *shallower* than the RAC slopes for the active stars with the same spectral types (i.e. dK4e, dK6e, and dM2e stars). As in H17, we have interpreted the *steeper* dependence on rotation in the active stars as an indication that, in such stars, an αΩ dynamo (with its greater sensitivity to rotation) is at work. On the other hand, the *shallower* slopes of the RAC in the inactive dK4, dK6, and dM2 stars is interpreted to mean that the dynamos in these stars are not as sensitive to rotation: specifically, we interpret the shallower slopes of the RAC as evidence that an αΩ dynamo is *not* at work in the inactive dK4, dK6, and dM2 stars. Instead, since an $\alpha^2$ dynamo is expected to be less sensitive to rotation, we believe that the shallow slopes of the RAC's in the inactive stars can be interpreted to mean that an $\alpha^2$ dynamo (or an $\alpha^2\Omega$ dynamo) is at work in those inactive stars.

When we examine the magnitudes of CaII emission flux in active dKe and dMe stars at these types (where F(αΩ) + F($\alpha^2$) plus a small acoustic component are present), and compare them to the fluxes in inactive dK and dM stars (where only F($\alpha^2$) plus a small acoustic component is present), we find (using e.g. Figs. 2, 3, and 5 in H17) that the mean CaII flux in the active stars exceeds that in the inactive stars by factors of a few. This suggests that, at least in K4-M2 dwarfs, it could be permissible to conclude that F(αΩ) exceeds F($\alpha^2$).

Let us hypothesize that at a certain spectral type, there is a transition from one magnetic mode to another: can we predict what we would see as we go to later spectral types? There are 3 possible outcomes: (a) The emission flux F(CaII) increases. (b) F(CaII) decreases. (c) F(CaII) stays the same. If we are to make a theoretical prediction between these possibilities, we need to have quantitative knowledge of the relative magnitudes of F(αΩ) and F($\alpha^2$).

The scenario in which we are most interested here is one in which F(αΩ) maintains a non-zero value down to a spectral type that is as late as Mx where the radiative core disappears. On the one hand, stars which are slightly hotter than stars of spectral type Mx (we label such stars as having spectral type Mx-) are expected to have an overall flux in CaII emission of F(a) + F(αΩ) + F($\alpha^2$): but because of the smallness of F(a), this is essentially equal to F(αΩ) + F($\alpha^2$). On the other hand, stars which are slightly cooler than stars with spectral type Mx (we label such stars as having spectral type Mx+) will have an overall flux in CaII emission of F(a) + F($\alpha^2$): once again, neglecting F(a) due to its smallness, this is essentially equal to F($\alpha^2$). Now let us consider two limiting cases: (i) F(αΩ) << F($\alpha^2$), and (ii) F(αΩ) ≥ F($\alpha^2$). In case (i), the disappearance of F(αΩ) will have a negligible effect on F(tot) at spectral type Mx. In such a case, therefore, there would be *no reason* to expect to see any observational signature in the empirical flux of CaII emission. It would indeed be unfortunate if we were unable to detect the location where F(αΩ) disappears.

On the other hand, in case (ii), at spectral type Mx-, F(tot) will be essentially equal to F(a) + F(αΩ), i.e. essentially F(αΩ), whereas at Mx+, F(tot) will be equal to F(a) + F($\alpha^2$), i.e. essentially F($\alpha^2$). Therefore F(tot) decreases essentially from F(αΩ) to F($\alpha^2$) when we pass through the transition from Mx- to Mx+. And since, by definition of case (ii) F(αΩ) is ≥ F($\alpha^2$), the emission flux in CaII across the Mx transition will have the following signature: a clear step *downward*.

The numerical value of the ratio of the two flux totals at Mx- and Mx+ is expected to be R(-+) = [F(a)+ F(αΩ) + F($\alpha^2$)]/ [F(a) + F($\alpha^2$)]. Essentially, R(-+) is equal to 1 + [F(αΩ)/ F($\alpha^2$)]. If it happens that F(αΩ) =

F($\alpha^2$), R(-+) would be found to have a value of 2. If it happens that F($\alpha\Omega$) > F($\alpha^2$), then R(-+) would be found to have a value in excess of 2. Therefore, if case (ii) is applicable, the "step downward" in flux level is expected to be by a factor of 2 or more.

This is reminiscent of the "dip" which can be seen in Fig. 3 above between spectral types M2.1 and M2.3: the step downward goes from (about) 1.1 at M- to (about) 0.5 at M+, i.e. an amplitude of 2.2. Similar steps, with amplitudes which are also about 2, also occur in Figs. 4 and 5 between the point labeled M1.8-M2.1 and the point labeled M2.2-M2.5. We suggest that the "dips" we have identified in Figs. 3, 4, and 5 may be a candidate for the "step downward" which was discussed above. If this is a correct interpretation, we interpret the range of M2.1-M2.3 as an improved estimate of the transition between an $\alpha\Omega$ dynamo and an $\alpha^2$ dynamo. In this context, the CaII line fluxes at spectral type M2.1 (and earlier) have access to an $\alpha\Omega$ dynamo as well as an $\alpha^2$ dynamo, whereas at spectral types M2.3 (and later), the stars can take advantage of only an $\alpha^2$ dynamo.

## 5. Conclusion

In this paper, we have reported on an expanded data set of CaII fluxes of chromospheric emission in a sample of roughly 600 M dwarfs. Our goal has been to determine if we can identify any signature in the chromospheric data which might be due to a transition from one form of dynamo operation to another. For example, the transition to complete convection (TTCC) which is traditionally predicted to occur in stars with masses of order 0.3-0.35 $M_\odot$ on the lower main sequence, is expected to lead to the suppression of an interface dynamo: the disappearance of such a dynamo might reasonably be expected to be accompanied by an observational signature of some kind in the strength of chromospheric emission. In an earlier search for such a signature, using a smaller sample (less than 300 stars: H17), we focused on determining the numerical values of the *slope* of the rotation-activity correlation (RAC) and used the slopes in search of a transition. Based on the behavior of the RAC slopes as a function of spectral type, H17 suggested that the **switch in dynamo mode** on the lower main sequence occurs at a spectral sub-type between M2 and M3. To the extent that this dynamo switch is associated with the TTCC, H17 suggested that the TTCC lies between M2 and M3.

About a year after H17 was published, a completely independent study (based on GAIA photometry: Jao et al 2018) has reported that a structural change in stellar structure, leading to a dip in the luminosity function, makes its appearance "near spectral type M3.0V". Theoretical modeling (MacDonald and Gizis 2018) suggests that the GAIA dip is associated with complicated evolutionary effects as a star approaches the TTCC: the complications lead to the temporary appearance of a *small convective core* in low mass stars in addition to the standard picture of radiative core plus outer convective envelope. It has long been known that dynamo activity in the Sun and stars is associated with the interface between radiative and convective regions: now, the possibility of a new (inner) interface emerges with the existence of a small (temporary) convective core, separated from the outer convective envelope by a radiative region. The new inner interface may add an additional source of dynamo activity in low-mass stars. As a result, the dynamo mode may not undergo a transition solely at "the" (traditional) TTCC, but also from the occurrence of a new dynamo mode deep in the interior.

With the expanded data set, we have, in this paper, examined the CaII emission fluxes in more detail. We have paid special attention to the range of spectral types from M1 to M3. We find that a "downward step" feature in this plot between spectral types M2.1 and M2.3 has properties which are consistent with the "switching off" of an important mode of chromospheric heating. We hypothesize that the mode of chromospheric heating which is switched off at the "step" is associated with an interface dynamo. We cannot say definitively whether the interface dynamo involved in the "downward step" lies at the inner or the outer interface, both of which are present during certain time intervals in the evolution of low-mass stars. If our hypothesis is correct, the new data suggest that the switch in dynamo mode may lie between M2.1 and M2.3. This range is consistent with, but more precise than, the conclusion of H17 that the switch in dynamo mode lies between M2 and M3. The new estimate of the location of the switch in dynamo mode corresponds to stars with T(eff) in the range 3610-3560 K.

We note that the location we have found for the switch in dynamo mode (M2.1-M2.3) is close to, but slightly earlier than, the spectral type "near M3.0V" which Jao et al (2018) have reported as the location of an empirical dip in the luminosity function. MacDonald and Gizis (2018) have suggested that the dip seen by Jao et al is associated with the transition to complete convection (TTCC) which is known to occur in lower main sequence stars (Limber 1958). Our data suggest that a switch in dynamo mode may occur at a spectral type which is close to, but (if we take our results literally) somewhat earlier, than the TTCC. We suggest that the switch in dynamo mode which we suggest occurs at M2.1-M2.3 may be associated with the appearance of a small convective core: this gives rise to a new (inner) interface between radiative and convective gas in the deepest interior of the star (in addition to the well-known outer interface between radiative core and outer convective envelope). **According to MacDonald and Gizis (2018), the dip seen by Jao et al (2018) occurs at the time when the two convective regions (core and envelope) merge into a single convective zone extending from center to surface**. MacDonald and Gizis have found that this merging process occurs sooner in lower mass M dwarfs, and later in more massive M stars: at least 9 b.y. in stars with mass 0.34 $M_\odot$, but less than 1 b.y. in stars with mass 0.31 $M_\odot$. In view of this difference in lifetimes, the inner interface dynamo is expected to survive longer in a more massive M dwarf, i.e. a star of earlier spectral type. This could bias the dynamo transition to somewhat earlier spectral types than the TTCC, perhaps explaining why we find M2.1-2.3 for the dynamo transition while the larger structural transition which appears in the luminosity function occurs at a slightly later spectral type, "near M3.0V" (Jao et al 2018).

**If we are correct in claiming that the presence of a second convective region (core) in an M dwarf may be associated with a switch in dynamo mode in such stars, then we need to consider that in other low-mass stars (including solar-like stars), a small convective core is also expected to be temporarily present. Could the presence of such a core lead to a signature of dynamo switching in solar-like stars also? We suggest two reasons why this is *not* likely to be detectable. First, in a solar-like star, the outer convective envelope remains so close to the surface (at radial locations of 0.7R(sun) and larger) that there is little or no opportunity for a merger with the small convective core. Second, the relative values of two time-scales are important: t(cc), the time interval during which the small convective core survives, and t(s), the ages of the stars in our sample. As MacDonald and Gizis (2018) have shown, stars with masses corresponding to early M spectral type undergo merging of central core and convective envelope on time-scales ranging from less than 1 Gyr to as long as 9 Gyr. Such time intervals are essentially identical with the (activity) ages of M dwarfs in the field (West et al 2008: especially their Fig. 10). Therefore, in our sample of 600 or so M dwarfs, it is highly likely that our stars include members with the correct age for merging of core and envelope. On the other hand, in a star with mass 1.0 M(sun), MacDonald (2020) reports that the convective core survives from age 29 Myr to**

**age 102 Myr, i.e. much shorter than the ages associated with M dwarfs. In view of this result, in order to detect any observational signatures of dynamo-switching which might be associated with such a core, we would have to examine activity data on solar-like stars with ages of 102 Myr and less. In a long-lasting observing program known as the *Sun in Time,* Guinan and Engle (2018) have reported on the coronal X-ray luminosity of solar-like stars (with spectral types G1.5V-G2.5V) with ages ranging up to as long as 7 Gyr. For present purposes, we are especially interested in the *youngest* stars in the Guinan and Engle sample: they turn out to be members of the Pleiades, which has an age of 130 Myr (Bell et al 2012). Thus, the data which are currently available for solar-like stars does *not* extend to ages which are young enough to overlap with the existence of a convective core. As a result, we do not yet have observations which would enable us to test whether or not an observational signature is present in the activity data related to the presence or absence of a small convective core.**

ACKNOWLEDGEMENTS

We thank the referee for extensive and constructive reports which have helped to sharpen the arguments in the paper. DJM thanks Dr J. MacDonald for a very helpful discussion of work on the inner convective core**, including numerical values for the duration of such a core in the Sun**, as well as for providing an estimate of T$_{eff}$ for stars near the TTCC.

**West, A. A., Hawley, S. L., Bochanski, J. J. et al. 2008, AJ 135, 785**

Withbroe, G. L., & Noyes, R. W. 1977, Ann. Rev. Astron. & Ap. 15, 363